\documentclass[11pt,a4paper]{article}
\usepackage{amssymb}

\usepackage{makeidx}
\usepackage{graphicx}
\usepackage{amsmath}

\newcounter{resultnum}[section]

\newcounter{conclusionnum}[section]

\newcounter{conditionnum}[section]

\newcounter{conjecturenum}[section]

\newcounter{examplenum}[section]

\newcounter{exercisenum}[section]

\newcounter{lemmanum}[section]

\newcounter{notationnum}[section]
\newtheorem{theorem}{Theorem}[section]

\newcounter{theoremnum}[section]
\newtheorem{definition}{Definition}[section]

\newcounter{definitionnum}[section]

\newcounter{corollarynum}[section]

\newcounter{remarknum}[section]

\newcounter{propositionnum}[section]

\newcounter{acknowledgementnum}[section]

\newcounter{algorithmnum}[section]

\newcounter{axiomnum}[section]

\newcounter{casenum}[section]

\newcounter{claimnum}[section]

\newcounter{summarynum}[section]

\newcounter{problemnum}[section]
\newenvironment{proof}[1][]{\textbf{Proof.} }{}

\newcommand{ \D} {\mbox{\rm I$\!$\bf{D}}}

\begin{document}

\title{ Clifford--Finsler Algebroids and \\
Nonholonomic Einstein--Dirac Structures}
\author{ Sergiu I. Vacaru\thanks{%
svacaru@brocku.ca } \\
-- \\
\textsl{\ Department of Mathematics, Brock University,}\\
\textsl{St. Catharines, Ontario, Canada L2S\ 3A1 }}
\date{March 5, 2006}
\maketitle

\begin{abstract}
We propose a new framework for constructing geometric and physical models on
nonholonomic manifolds provided both with Clifford -- Lie algebroid symmetry
and nonlinear connection structure. Explicit parametrizations of generic
off--diagonal metrics and linear and nonlinear connections define different
types of Finsler, Lagrange and/or Riemann--Cartan spaces. A generalization
to spinor fields and Dirac operators on nonholonomic manifolds motivates the
theory of Clifford algebroids defined as Clifford bundles, in general,
enabled with nonintegrable distributions defining the nonlinear connection.
In this work, we elaborate the algebroid spinor differential geometry and
formulate the (scalar, Proca, graviton, spinor and gauge) field equations on
Lie algebroids. The paper communicates new developments in geometrical
formulation of physical theories and this approach is grounded on a number
of previous examples  when exact solutions with generic off--diagonal
metrics and generalized symmetries in modern gravity define nonholonomic
spacetime manifolds with uncompactified extra dimensions.

\vskip0.3cm \textbf{Keywords:}\ Lie algebroids, Clifford algebroids, Finsler
and Lagrange geometry, exact solutions, Einstein--Dirac equations, string
and Einstein gravity.

\vskip0.2cm

2000 AMS Subject Classification:\

15A66, 17B99, 53A40, 58B20, 81R25, 83C20, 83C60, 83E99
\end{abstract}




\section{Introduction}

A class of spacetimes possessing noncommutative and/or Lie algebroid
symmetries can be defined as exact solutions in string and Einstein gravity %
\cite{vncsym,valgexsol}. This leads to new developments in  formulation of
classical and quantum field theories following the  geometry of nonholonomic
manifolds \cite{esv} possessing Lie algebroid  symmetry \cite{acsw}. Lie
algebroid structures in gravity are modelled by generic off--diagonal
metrics and nonholonomic frames (vielbeins)  with associated nontrivial
nonlinear connection (N--connection).  The spacetimes provided with
compatible metric, linear connection  and N--connection structures and
possessing Lie algebroid  symmetry are called Einstein--Cartan algebroids,
or (in a more  general context, for various extensions of the
Riemann--Cartan geometry) Lie N--algebroids. Usually, the Lie algebroids can
be defined for a vector, or tangent, bundle but, in general, they can be
considered for any nonholonomic manifold provided with a nonintegrable
(nonholonomic) distribution\footnote{%
In our works we use distributions defining N--connection  structures with
the coefficients induced by the metric's off--diagonal terms and
corresponding vielbein's coefficients. The geometric constructions are
performed for nonholonomic manifolds, i. e. spaces provided with
nonintegrable distributions. In a particular case, when such distributions
are related to the exact sequences of subspaces defining a N--connection,
the spaces are called N--anholonomic.}. In brief, such spaces are called Lie
N--algebroids. Similar constructions elaborated for the Einstein--Dirac
spaces give rise to the geometry of Clifford algebroids. If the curved
spinor spaces are also enabled with  Finsler, or Lagrange, structures, we
deal with Clifford--Finsler, or\ Clifford--Lagrange, algebroids.

We note that the methods of Finsler and Lagrange geometry \cite{ma,v2} were
recently reconsidered in a new way in order to solve physical problems
related to standard theories of gravity and field interactions \cite%
{vncsym,vncl,v2,vt,vp}. If the former physical applications of Finsler
geometry were elaborated  on tangent/ vector bundles, with less
straightforward connections to standard physical theories (see reviews and
references in \cite{vncsym,vncl,v2}), in our approach we tried to define
Finsler like structures as exact solutions in Einstein and extra dimension
gravity \cite{vt,vp} when certain dimensions  are not compactified. Such
constructions are related to the geometry of nonholonomic manifolds
possessing generalized symmetries (Lie algebroid and/or Clifford symmetries,
noncommutative structures induces by anholonomic frames, ...) and a number
of ideas and methods from Finsler geometry seem to be of general interest
and significant importance for physical applications. Here we note, that
this paper in not just on Clifford--Finsler geometry and related Lie
algebroid structures but rather on  (pseudo) Riemann geometry and
gravitational and field  interactions (and extensions to non--trivial
torsion induced, for  instance, from string theory and/or by nonholonomic
frame  effects)  when the spinor and Lie aglebroid structures are defined in
nonholonomic  form and certain methods from Finsler geometry became very
important  and efficient in order to solve nonlinear physical problems.

This work develops the geometry of Clifford N--algebroids and generalized
Finsler--spinor spaces elaborated in Refs. \cite{vjmp,vhsp,vstav,vncl}.  If
the first applications of algebroid methods were in geometric mechanics \cite%
{weins1,lib,mart,dl1}, the recent works suggest a very promising route
toward the theory of gauge fields, gravity and strings and noncommutative
geometry \cite{vncl,vncgg,vncsym}. We cite Ref. \cite{acsw} for details on
algebroid theory and related bibliography.

In the present paper, we address essentially the following two purposes:\
\ The \textbf{first one} is  to define and study the geometry
of Clifford algebroids and their
N--anholonomic deformations, Clifford N--algebroids, and analysis of theirs
main properties in relation to spinors in gravity theories and on
nonholonomic manifolds. The \textbf{second} aim is the formulation of the
field equations on Lie algebroids.

The structure of the paper is the following:\
The theory of Clifford algebroids is formulated in section 2: we
remember the main definitions of nonholonomic manifolds provided with
N--connection structure, define Clifford
N--algebroids and study the related spinor differential geometry. Section 3
is devoted to the field equations on N--anholonomic manifolds and their
redefinition on Clifford N--anholonomic algebroids. We start with a study of
the Dirac operator and spin connections on nonholonomic manifolds. Then the
constructions are completed with spinor formulations of the basic equations
for scalar, Proca, graviton, Dirac and gauge fields interactions and related
Lie/Clifford N--algebroid structures. In section 4, we present conclusions
and outlook.

\section{Clifford Algebroids and N--Connections}
The geometry of spinor spaces enabled with nonlinear connection
(N--con\-nec\-ti\-on) struc\-ture was elaborated in a series of works \cite%
{vjmp,vhsp,vstav} (see also papers \cite{penr1,penr2,lue} for general
references on Clifford and spinor differential geometry and applications to
physics). Here we note that the concept of N--connection was originally
proposed in the framework of Finsler geometry and geometric mechanics but
such nonholonomic structures \footnote{%
defined by exact sequences of subspaces of the tangent space to the
spacetime manifold and related nonintegrable distributions} may be also
considered on (pseudo) Riemannian and Einstein--Cartan--Weyl spaces, see
discussion and historical remarks in Ref. \cite{vncl}. A class of
nonholonomic spinor configurations can be defined by exact solutions of the
Einstein--Dirac equations parametrized by generic off--diagonal metric
ansatz, nonholonomic vielbeins associated to nontrivial N--connections and
arbitrary linear connections with nontrivial torsion.

The aim of this section is to formulate the theory of Clifford algebroids
provided with nonlinear connection  (N--connection) structures, i. e. the
theory of Clifford N--algebroids. For holonomic configurations, the Clifford
algebroids can be defined as usual Lie algebroids \cite{acsw} but associated
to a Clifford bundle instead to a vector or tangent bundle.

\subsection{Nonholonomic manifolds and nonlinear conections}

We outline some basic definitions and formulas from  the geometry of
manifolds provided with N--connection structure, see details in Refs. \cite%
{v2,vncl}.

Let us consider a Riemann--Cartan manifold $V$ of dimension $n+m$ and
necessary smooth class and provided with general metric (of arbitrary
signature) and linear connection structures. The local coordinates are
denoted $u=(x,y), $ or $u^{\alpha }=\left( x^{i},y^{a}\right) ,$ where the
abstract, or coordinate, indices take respectively the values $%
i,j,k,...=1,2,...,n$ and $a,b,c,...=n+1,n+2,...,n+m.$ Such a splitting of
dimension and coordinates will be adapted bellow to the nonlinear connection
structure. We denote by $M$ a subspace of $V,$ $\dim M=n,$ provided with
local coordinates $x^{i}.$ The metric on $V$ is parametrized in the form
\begin{equation}
\mathbf{g}=g_{\alpha \beta }\mathbf{e}^{\alpha }\otimes \mathbf{e}^{\beta
}=g_{ij}(u)\mathbf{e}^{i}\otimes \mathbf{e}^{j}+h_{ab}(u)\mathbf{e}%
^{a}\otimes \mathbf{e}^{b}  \label{metr}
\end{equation}%
where
\begin{equation}
\mathbf{e}^{\mu }=[e^{i}=dx^{i},\mathbf{e}^{a}=dy^{a}+N_{i}^{a}(u)dx^{i}]
\label{ddif}
\end{equation}%
is the dual frame to
\begin{equation}
\mathbf{e}_{\nu }=[\mathbf{e}_{i}=\frac{\partial }{\partial x^{i}}%
-N_{i}^{a}(u)\frac{\partial }{\partial y^{a}},e_{a}=\frac{\partial }{%
\partial y^{a}}].  \label{dder}
\end{equation}%
Such vielbeins are called N--adapted frames. \footnote{%
In order to preserve a relation with the previous denotations \cite%
{vncl,vjmp,vhsp,vstav}, we note that $\mathbf{e}_{\nu }=(e_{i},e_{a})$ and $%
\mathbf{e}^{\mu }=(e^{i},e^{a})$ are, respectively, the former $\delta _{\nu
}=\delta /\partial u^{\nu }=(\delta _{i},\partial _{a})$ and $\delta ^{\mu
}=\delta u^{\mu }=(d^{i},\delta ^{a})$ which emphasize that operators (\ref%
{dder}) and (\ref{ddif}) define, correspondingly, certain ``N--elongated''
partial derivatives and differentials which are more convenient for
calculations on such nonholonomic manifolds.}

We denote by $\pi ^{\top }:TV\rightarrow TM$ the differential of a map $\pi
:TV\rightarrow hV,$ where $hV$ is locally isomorphic to $M,$ defined by
fiber preserving morphisms of the tangent bundles $TV$ and $TM.$ The kernel
of $\pi ^{\top }$is just the vertical subspace $vV,\ \dim (vV)=m,$ with a
related inclusion mapping $i:vV\rightarrow TV$ and $hV$ is a horizontal
subspace. It should be emphasized that one exists such maps and local
decompositions when $V\rightarrow M$ is a surjective submersion. A
particular case is that of a fiber bundle but we can obtain the results in
the general case. \footnote{%
see discussions and references in \cite{esv,vncl} related to almost
sympletic manifolds, (pseudo) Riemannian spaces and vector bundles and
generalizations} \ A \textbf{nonlinear connection (N--connection) }$\mathbf{N%
}$ on a manifold $V$ is defined by the splitting on the left of an exact
sequence
\begin{equation*}
0\rightarrow vV\rightarrow TV\rightarrow TV/vV\rightarrow 0,
\end{equation*}%
i. e. by a morphism of submanifolds $\mathbf{N:\ \ }TV\rightarrow vV$ such
that $\mathbf{N\circ i}$ is the unity in $vV.$

Equivalently, a N--connection is defined by a Whitney sum of horizontal (h)
subspace, $hV\simeq M$ (we shall use the symbol ''$\simeq $'' in order to
emphasize some isomorphisms of spaces) and vertical (v) subspace, $vV,$
\begin{equation}
T\mathbf{V}=hV\oplus vV.  \label{whitney}
\end{equation}%
The spaces provided with N--connection structure are denoted by boldface
symbols. For instance, we write $\mathbf{V}$ for a manifold $V$ provided
with a distribution (\ref{whitney}) (being, in general, nonintegrable, i. e.
nonholonomic\footnote{%
in literature it is also used an equivalent term:\ anholonomic}). \ Such
manifolds are called N--anholonomic with the nonholonomy defined by a
N--connection structure. In a similar manner, we can define nonholonomic
manifolds enabled with certain more general nonintegrable (nonholonomic)
distributions of subspaces in $TV,$ or in $TTV,$ and so on ... but in this
paper we shall restrict our considerations only to N--anholonomic manifolds
with N--connection splitting on $TV.$ We shall use boldfaced indices for the
geometric objects adapted to the N--connection.

\ Locally, a N--connection is defined by its coefficients $N_{i}^{a}(u),$%
\begin{equation*}
\mathbf{N}=N_{i}^{a}(u)dx^{i}\otimes \frac{\partial }{\partial y^{a}}.
\end{equation*}%
The well known class of linear connections consists on a particular subclass
with the coefficients being linear on $y^{a},$ i.e., $N_{i}^{a}(u)=\Gamma
_{bj}^{a}(x)y^{b}.$ Any N--connection is characterized by its N--connection
curvature
\begin{equation*}
\mathbf{\Omega }=\frac{1}{2}\Omega _{ij}^{a}dx^{i}\wedge dx^{j}\otimes \frac{%
\partial }{\partial y^{a}},
\end{equation*}%
with N--connection curvature coefficients%
\begin{equation*}
\Omega _{ij}^{a}=\delta _{\lbrack j}N_{i]}^{a}=\delta _{j}N_{i}^{a}-\delta
_{i}N_{j}^{a}=\frac{\partial N_{i}^{a}}{\partial x^{j}}-\frac{\partial
N_{j}^{a}}{\partial x^{i}}+N_{i}^{b}\frac{\partial N_{j}^{a}}{\partial y^{b}}%
-N_{j}^{b}\frac{\partial N_{i}^{a}}{\partial y^{b}},
\end{equation*}%
and states the condition that the vielbeins (\ref{ddif}) satisfy the
nonholonomy (equivalently, anholonomy) relations
\begin{equation*}
\lbrack \mathbf{e}_{\alpha },\mathbf{e}_{\beta }]=\mathbf{e}_{\alpha }%
\mathbf{e}_{\beta }-\mathbf{e}_{\beta }\mathbf{e}_{\alpha }=W_{\alpha \beta
}^{\gamma }\mathbf{e}_{\gamma }
\end{equation*}%
with (antisymmetric) nontrivial anholonomy coefficients $W_{ia}^{b}=\partial
_{a}N_{i}^{b}$ and $W_{ji}^{a}=\Omega _{ij}^{a}.$

All our further geometric constructions will be for spaces with nonholonomic
splitting (\ref{whitney}) and performed in 'N--adapted' form with respect to
local frames of type  (\ref{ddif}) and  (\ref{dder}).

\subsection{Clifford N--algebroids}

Let us state the notations for abstract (coordinate) d--tensor indices of
geometrical objects defined with respect to an arbitrary (coordinate) local
basis, i. e. system of reference. For a local basis on $\mathbf{V,}$ we
write $e_{\alpha }=(e_{i},v_{a}).$ The small Greek indices $\alpha ,\beta
,\gamma ,...$ are considered to be general ones, running values $1,2,\ldots
,n+m$ and $i,j,k,...$ and $a,b,c,...$ respectively label the geometrical
objects on the base and typical ''fiber'' and run, correspondingly, the
values $1,2,...,n$ and $1,2,...,m.$ The dual base is denoted by $e^{\alpha
}=(e^{i},v^{a}).$ The local coordinates of a point $u\in \mathbf{V}$ are
written $\mathbf{u=}(x,y),\ $or $u^{\alpha }=(x^{i},y^{a}),$ where $y^{a}$
is the $a$--th coordinate with respect to the basis $(v_{a})$ and $(x^{i})$
are local coordinates on $h\mathbf{V}$ with respect to $e_{i}\mathbf{.}$ We
shall use ''boldface'' symbols in order to emphasize that the objects are
defined on spaces provided with N--connection structure.

We suppose that the N--anholonomic manifold $\mathbf{V}$ admits a d--spinor
structure which allows us to introduce spinor coordinates and
parametrizations of geometrical objects.
 Let%
\begin{equation*}
\mathbf{e}_{\acute{\alpha}}^{\quad \mathbf{\acute{\alpha}}}=\left( e_{\acute{%
\imath}}^{\quad \mathbf{\acute{1}}},e_{\acute{\imath}}^{\quad \mathbf{\acute{%
2}}},...,e_{\acute{\imath}}^{\quad \mathbf{\tilde{k}}_{(n)}},e_{\acute{a}%
}^{\quad \mathbf{\acute{1}}},e_{\acute{a}}^{\quad \mathbf{\acute{2}}},...,e_{%
\acute{a}}^{\quad \mathbf{\tilde{k}}_{(m)}}\right) ,
\end{equation*}%
with boldfaced indices running coordinate values
on dimensions of d--spinor spaces, $\mathbf{\tilde{k}}_{(n)}$ and $\mathbf{%
\tilde{k}}_{(m)},$  be the coefficients of a
d--spinor basis
\begin{equation}
\mathbf{e}_{\acute{\alpha}}=(e_{\acute{\imath}},e_{\acute{a}}).  \label{dsb}
\end{equation}%
The dual basis (co--basis)
\begin{equation}
\mathbf{e}^{\acute{\alpha}}=(e^{\acute{\imath}},e^{\acute{a}})  \label{cdsb}
\end{equation}
has the coefficients
\begin{equation*}
\mathbf{e}_{\mathbf{\acute{\alpha}}}^{\quad \acute{\alpha}}=\left( e_{%
\mathbf{\acute{1}}}^{\quad \acute{\imath}},e_{\mathbf{\acute{2}}}^{\quad
\acute{\imath}},...,e_{\mathbf{\tilde{k}}_{(n)}}^{\quad \acute{\imath}},e_{%
\mathbf{\acute{1}}}^{\quad \acute{a}},e_{\mathbf{\acute{2}}}^{\quad \acute{a}%
},...,e_{\mathbf{\tilde{k}}_{(m)}}^{\quad \acute{a}}\right).
\end{equation*}%
Similar formulas hold for the associated d--spinor spaces provided with
local bases $\mathbf{e}_{\mathbf{\grave{\alpha}}}^{\quad \grave{\alpha}}$
and $\mathbf{e}_{\grave{\alpha}}^{\quad \mathbf{\grave{\alpha}}}.$ Such
spinor bases are stated to be compatible to the N--connection splitting, i.
e. to the vielbeins (\ref{dder}) and (\ref{ddif}). For a given d--metric
structure on $\mathbf{V}$ and its spinor decomposition,
 with associated spinor bases $%
\mathbf{e}_{\grave{\alpha}}=(e_{\grave{\imath}},e_{\grave{a}})$, which
allows us to introduce the $\gamma $--objects, we can define, for instance,
a N--adapted basis
\begin{equation*}
\mathbf{e}_{\alpha }=\left( \gamma _{\alpha }\right) ^{\acute{\alpha}\grave{%
\alpha}}\mathbf{e}_{\acute{\alpha}}\mathbf{e}_{\grave{\alpha}}=\left[
e_{i}=\left( \gamma _{i}\right) ^{\acute{\imath}\grave{\imath}}e_{\acute{%
\imath}}e_{\grave{\imath}},e_{a}=\left( \gamma _{a}\right) ^{\acute{a}%
\grave{a}}e_{\acute{a}}e_{\grave{a}}\right] .
\end{equation*}%
As a result, we can alternatively consider spinor coordinates, for instance,
\begin{equation*}
u^{\alpha }=(x^{i},y^{a})\rightarrow u^{\acute{\alpha}\grave{\alpha}}=(x^{%
\acute{\imath}\grave{\imath}},y^{\acute{a}\grave{a}}).
\end{equation*}%
For even dimensions of $n,$ or $m,$ further reductions are possible, when $%
x^{\acute{\imath}\grave{\imath}}\rightarrow x^{II^{\prime }},$ or $y^{%
\acute{a}\grave{a}}\rightarrow y^{AA^{\prime }}.$ This way, the d--tensor
indices can be transformed into the d--spinor ones and inversely.

The standard definition of a Lie algebroid $\mathcal{A}\doteqdot (E,\left[
\cdot ,\cdot \right] ,\rho )$ is associated a vector bundle $\mathcal{E}%
=(E,\pi ,M),$ with a surjective map $\pi :E\longrightarrow M$ of the total
spaces $E$ to the base manifold $M,$ of respective dimensions $\dim E=n+m$
and $\dim M=n.$ The algebroid structure is stated by the anchor map $\rho :\
E\rightarrow TM$ \ ($TM$ is the tangent bundle to $M$) and a Lie bracket on
the $C^{\infty }(M)$--module of sections of $E,$ denoted $Sec(E),$ such that
\begin{equation*}
\left[ X,fY\right] =f\left[ X,Y\right] +\rho (X)(f)Y
\end{equation*}%
for any $X,Y\in Sec(E)$ and $f\in C^{\infty }(M).$ The anchor also induces a
homomorphism of $C^{\infty }(M)$-modules $\rho :Sec(A)\rightarrow \mathcal{X}%
^{1}(M)$ where $\wedge ^{r}(M)$ and $\mathcal{X}^{r}(M)$ will denote,
respectively, the spaces of differential $r$--forms and $r$--multivector
fields on $M.$

In local form, the Lie algebroid structure on the manifold $\mathbf{V}$ is
defined by its structure functions $\rho _{a}^{i}(x)$ and $C_{ab}^{f}(x)$
defining the relations
\begin{eqnarray}
\ \rho (e_{a}) &=&\rho _{a}^{i}(x)\ e_{i}=\rho _{a}^{\underline{i}}(x)\
\partial _{\underline{i}},  \label{anch} \\
\lbrack e_{a},e_{b}] &=&C_{ab}^{c}(x)\ e_{c}  \label{liea}
\end{eqnarray}%
and subjected to the structure equations
\begin{equation}
\rho _{a}^{j}\frac{\partial \rho _{b}^{i}}{\partial x^{j}}-\rho _{b}^{j}%
\frac{\partial \rho _{a}^{i}}{\partial x^{j}}=\rho _{c}^{j}C_{ab}^{c}~%
\mbox{\ and\ }\sum\limits_{cyclic(a,b,c)}\left( \rho _{a}^{j}\frac{\partial
C_{bc}^{d}}{\partial x^{j}}+C_{af}^{d}C_{bc}^{f}\right) =0;  \label{lasa}
\end{equation}%
for simplicity, we shall omit underlying of coordinate indices if it will
not result in ambiguities. Such equations are standard ones for the Lie
algebroids but defined on a N--anholonomic manifolds. In brief, we call them
Lie N--algebroids.

\begin{definition}
A Clifford algebroid $\mathcal{C}(E)\doteqdot (\mathbb{C}l(E),\ ^{s}\left[
\cdot ,\cdot \right] ,\ ^{s}\rho )$ $\ $is associated to a Clifford bundle $%
\mathbb{C}l(E)\doteq \mathbb{C}l(T^{\ast }E)$ defined by the vector bundle $%
\mathcal{E}=(E,\pi ,M)$ and provided with ''spin'' anchor $^{s}\rho $ and
(Lie type) commutator structure $\ ^{s}\left[ \cdot ,\cdot \right] $ \
defined on the $\ $Clifford module $\ Sec(\mathbb{C}l(M)).$
\end{definition}

The Clifford algebroid strucure on a manifold $M$ is defined $\mathcal{C}%
(TM)\doteqdot (\mathbb{C}l(TM),\ ^{s}\left[ \cdot ,\cdot \right] ,\ ^{s}\rho
).$

In local form, the spinor structure functions are written
\begin{eqnarray}
\ \rho (e_{\acute{a}\grave{a}}) &=&\rho _{\acute{a}\grave{a}}^{i}(x)\
e_{i}=\rho _{\acute{a}\grave{a}}^{\underline{i}}(x)\ \partial _{\underline{i}%
},  \label{anchs} \\
\lbrack e_{\acute{a}\grave{a}},e_{\acute{b}\grave{b}}] &=&C_{\acute{a}%
\grave{a}\acute{b}\grave{b}}^{\acute{c}\grave{c}}(x)\ e_{\acute{c}\grave{c}},
\label{lieas}
\end{eqnarray}%
where we can consider a spinor decomposition on $M$ with re--definition of
indices like $i\rightarrow \acute{\imath},\grave{\imath}.$ Such structure
functions can be induced by pure spinor ones,
\begin{equation}
\ \rho (e_{\acute{a}})=\rho _{\acute{a}}^{i}(x)\ e_{i}  \label{anchps}
\end{equation}%
and
\begin{equation*}
\lbrack e_{\acute{a}},e_{\acute{b}}]=C_{\acute{a}\acute{b}}^{\acute{c}}(x)\
e_{\acute{c}},
\end{equation*}%
where, for instance, we can consider $\rho _{\acute{a}\grave{a}}^{i}(x)=\rho
_{\acute{a}}^{i}(x)\rho _{\grave{a}}^{i}(x)$ for any fixed value of $i.$ The
structure equations (\ref{lasa}) can be written in spinor form by
introducing spinor variables (see examples of calculus with spinors in the
next section).

We can write down the Lie algebroid and N--connection structures in a
compatible form by introducing the ''N--adapted'' anchor%
\begin{equation}
{}\widehat{\mathbf{\rho }}_{a}^{j}(x,u)\doteqdot \mathbf{e}_{\ \underline{j}%
}^{j}(x,u)\mathbf{e}_{a}^{\ \underline{a}}(x,u)\ \rho _{\underline{a}}^{%
\underline{j}}(x)  \label{bfanch}
\end{equation}%
and ''N--adapted'' (boldfaced) structure functions
\begin{equation}
\mathbf{C}_{ag}^{f}(x,u)=\mathbf{e}_{\ \underline{f}}^{f}(x,u)\mathbf{e}%
_{a}^{\ \underline{a}}(x,u)\mathbf{e}_{g}^{\ \underline{g}}(x,u)\ C_{%
\underline{a}\underline{g}}^{\underline{f}}(x),  \label{bfstrf}
\end{equation}%
respectively, into formulas (\ref{anch}), (\ref{liea}) and (\ref{lasa}). In
general, the RC--algebroids are defined by the corresponding sets of
functions ${}\widehat{\mathbf{\rho }}_{a}^{j}(x,y)$ and $\mathbf{C}%
_{ag}^{f}(x,y)$ with additional dependencies on v--variables $y^{b}$ for the
N--adapted structure functions. For such Lie N--algebroids, the structure
relations became
\begin{eqnarray}
{}\widehat{\mathbf{\rho }}(e_{b}) &=&{}\widehat{\mathbf{\rho }}%
_{b}^{i}(x,y)\ e_{i},  \label{anch1d} \\
\lbrack e_{d},e_{b}] &=&\mathbf{C}_{db}^{f}(x,y)\ e_{f}  \label{lie1d}
\end{eqnarray}%
and the structure equations of the Lie N--algebroid are written
\begin{eqnarray}
{}\widehat{\mathbf{\rho }}_{a}^{j}e_{j}({}\widehat{\mathbf{\rho }}%
_{b}^{i})-{}\widehat{\mathbf{\rho }}_{b}^{j}e_{j}({}\widehat{\mathbf{\rho }}%
_{a}^{i}) &=&{}\widehat{\mathbf{\rho }}_{e}^{j}\mathbf{C}_{ab}^{e},
\label{lased} \\
\sum\limits_{cyclic(a,b,e)}\left( {}\widehat{\mathbf{\rho }}_{a}^{j}e_{j}(%
\mathbf{C}_{be}^{f})+\mathbf{C}_{ag}^{f}\mathbf{C}_{be}^{g}-\mathbf{C}%
_{b^{\prime }e^{\prime }}^{f^{\prime }}{}\widehat{\mathbf{\rho }}_{a}^{j}%
\mathbf{Q}_{f^{\prime }bej}^{fb^{\prime }e^{\prime }}\right) &=&0,  \notag
\end{eqnarray}%
for $\mathbf{Q}_{f^{\prime }bej}^{fb^{\prime }e^{\prime }}=\mathbf{e}_{\
\underline{b}}^{b^{\prime }}\mathbf{e}_{\ \underline{e}}^{e^{\prime }}%
\mathbf{e}_{f^{\prime }}^{\ \underline{f}}~e_{j}(\mathbf{e}_{b}^{\
\underline{b}}\mathbf{e}_{e}^{\ \underline{e}}\mathbf{e}_{\ \underline{f}%
}^{f})$ with the values $\mathbf{e}_{\ \underline{b}}^{b^{\prime }}$ and $%
\mathbf{e}_{f^{\prime }}^{\ \underline{f}}$ defined by the \ N--connection.
The Lie N--algebroid structure will be characterized by the data ${}\widehat{%
\mathbf{\rho }}_{b}^{i}(x,y)$ and $\mathbf{C}_{db}^{f}(x,y)$ stated with
respect to the N--adapted frames (\ref{dder}) \ and (\ref{ddif}).

A Riemann--Cartan algebroid (in brief, RC--algebroid) is a Lie algebroid $\
\mathcal{A}\doteqdot (\mathbf{V},\left[ \cdot ,\cdot \right] ,\rho )$
associated to a N--anholonomic manifold $\mathbf{V}$ provided with a
N--connection $\mathbf{N},$ symmetric metric $\mathbf{g(u)}$ and linear
connection $\mathbf{\Gamma (u)}$ structures resulting in a metric compatible
and N--adapted covariant derivative $\mathbf{D},$ when $\mathbf{Dg=0,}$ but,
in general, with nonvanishing torsion. In spinor variables, the
RC--algebroids transform into Clifford N--algebroids associated to
corresponding N--anholonomic manifolds instead of vector bundles. They are
characterized by the same set of relations (\ref{bfanch})--(\ref{lased})
re--written in d--spinor variables.

\subsection{N--algebroid spinor differential geometry}

The goal of the section is to outline the main results from the differential
geometry of d--spinors for \ the Clifford N--algebroids and related
N--anholonomic manifolds. The d--tensor and d--connection formulas and basic
equations are investigated in details in Ref. \cite{valgexsol}. Such Lie
N--algebroid relations can be obtained by ''anchoring'' the formulas for
d--connections, d--torsions and d--curvatures stated.
 In result, one obtains certain differential geometric objects on the set
of sections like $Sec(v\mathbf{V})$ or $Sec(\mathbf{E}),$ when the ''fiber''
derivatives are changed into horizontal ones, $\partial /\partial
y^{a}\rightarrow \rho _{a}^{i}\partial /\partial x^{a},$ or in N--adapted
form, $e_{a}\rightarrow {}\widehat{\mathbf{\rho }}_{a}^{j}e_{j}.$ In spinor/
d--spinor variables, such formulas transform into certain analogous on
Clifford N--algebroids provided with arbitrary but N--adapted and compatible
d--metric and d-- connection structure.

We use denotations
\begin{equation*}
e^{\alpha }=(e^{i},e^{a})\in \mathcal{\gamma }{^{\alpha }}=(\mathcal{\gamma }%
{^{i},\mathcal{\gamma }^{a}})\,\mbox{ and }\zeta ^{\acute{\alpha}}=(\zeta ^{%
\acute{\imath}},\zeta ^{\acute{a}})\in \mathcal{\gamma }{^{\acute{\alpha}}}=(%
\mathcal{\gamma }{^{\acute{\imath}},\mathcal{\gamma }^{\acute{a}}})\,
\end{equation*}%
for, respectively, elements of modules of d-vector and irreduced d--spinor
fields (see details in \cite{vjmp}). D-tensors and
d--spinor tensors (irreduced or reduced) will be interpreted as elements of
corresponding $\mathcal{\gamma }$--modules, for instance,
\begin{equation*}
q_{~\beta ...}^{\alpha }\in \mathcal{\gamma ^{\alpha }{}_{\beta }},\psi _{~%
\acute{\beta}\quad ...}^{\acute{\alpha}\quad \acute{\gamma}}\in \mathcal{%
\gamma }_{~\acute{\beta}\quad ...}^{\acute{\alpha}\quad \acute{\gamma}}~,\xi
_{\quad JK^{\prime }N^{\prime }}^{II^{\prime }}\in \mathcal{\gamma }_{\quad
JK^{\prime }N^{\prime }}^{II^{\prime }}~,...
\end{equation*}

We can establish a correspondence between the d--metric $g_{\alpha \beta }$ (%
\ref{metr}) and d--spinor metric $\epsilon _{\acute{\alpha}\acute{\beta}}$
 for both h- and v-subspaces of $\mathbf{V)}$  by using the relation
\begin{equation}
g_{\alpha \beta }=\frac{1}{\tilde{k}_{(n)}+\tilde{k}_{(m)}}((\gamma
_{(\alpha }(u))^{\acute{\alpha}\grave{\alpha}}(\gamma _{\beta )}(u))^{\acute{%
\beta}\grave{\beta}})\epsilon _{\acute{\alpha}\grave{\beta}}\epsilon _{%
\acute{\beta}\grave{\alpha}},  \label{2.78}
\end{equation}%
where $(\alpha \beta )$ denotes symmetrization on such indices and
\begin{equation}
(\gamma _{\alpha }(u))^{\acute{\alpha}\grave{\alpha}}=l_{\alpha }^{\widehat{%
\alpha }}(u)(\gamma _{\widehat{\alpha }})^{\acute{\alpha}\grave{\alpha}}.
\label{2.79}
\end{equation}%
 In brief, we
can write (\ref{2.78}) in the form
\begin{equation}
g_{\alpha \beta }=\epsilon _{\acute{\alpha}\grave{\beta}}\epsilon _{\acute{%
\beta}\grave{\alpha}}  \label{2.80}
\end{equation}%
if the $\gamma $-objects are considered as a fixed structure, whereas $%
\epsilon $-objects are treated as caring the metric ''dynamics ''. This
variant is used, for instance, in the so-called 2-spinor geometry \cite%
{penr1,penr2} and should be preferred if we have to make explicit the
algebraic symmetry properties of d--spinor objects. An alternative way is to
consider as fixed the algebraic structure of $\epsilon $-objects and to use
variable components of $\gamma $-objects of type (\ref{2.79}) for developing
a variational d--spinor approach to gravitational and matter field
interactions (the spinor Ashtekar variables \cite{ash} are introduced in
this manner). In this paper we shall follow in the bulk the first approach
but we note that the second type of spinor calculus is more convenient for
finding exact solutions with nonholonomic variables.

We note that a d--spinor metric
\begin{equation*}
\epsilon _{\acute{\alpha}\grave{\beta}}=\left(
\begin{array}{cc}
\epsilon _{\acute{\imath}\grave{j}} & 0 \\
0 & \epsilon _{\acute{a}\grave{b}}%
\end{array}%
\right)
\end{equation*}%
on the d--spinor space $\mathbf{S}=(S_{(h)},S_{(v)})$ may have symmetric or
antisymmetric h (v) -components $\epsilon _{\acute{\imath}\grave{j}}$ ($%
\epsilon _{\acute{a}\grave{b}}).$  For
simplicity, in this section (in order to avoid cumbersome calculations
connected with eight-fold periodicity on dimensions $n$ and $m$ on a
N--anholonomic manifold) we shall develop a general d--spinor formalism only
by using irreduced spinor spaces $\mathbf{S}_{(h)}$ and $\mathbf{S}%
_{(h)}^{\prime }.$

\subsubsection{ D--covariant derivation}

For a d--covariant operator
\begin{equation*}
\mathbf{D}_{\alpha }=\left( D_{i},D_{a}\right) =(\gamma _{\alpha })^{\acute{%
\alpha}\grave{\alpha}}\mathbf{D}_{\acute{\alpha}\grave{\alpha}}=\left(
(\gamma _{i})^{\acute{\imath}\grave{\imath}}\ D_{\acute{\imath}\grave{\imath}%
},~(\gamma _{a})^{\acute{a}\grave{a}}\ D_{\acute{a}\grave{a}}\right)
\end{equation*}%
(in brief, we shall write
\begin{equation*}
\mathbf{D}_{\alpha }=\mathbf{D}_{\acute{\alpha}\grave{\alpha}}=\left( D_{%
\acute{\imath}\grave{\imath}},~D_{\acute{a}\grave{a}}\right) ),
\end{equation*}%
being constructed by using the coefficients of a d--connection, we define
the action on a d--spinor $\mathcal{\gamma }^{\acute{\beta}}$ as a map
\begin{equation*}
\mathbf{D}_{\acute{\alpha}\grave{\alpha}}\ :\ \mathcal{\gamma }^{\acute{\beta%
}}\rightarrow \gamma _{\alpha }^{\acute{\beta}}=\gamma _{\acute{\alpha}%
\grave{\alpha}}^{\acute{\beta}}
\end{equation*}%
satisfying conditions
\begin{equation*}
\mathbf{D}_{\alpha }(\xi ^{\acute{\beta}}+\eta ^{\acute{\beta}})=\mathbf{D}%
_{\alpha }\xi ^{\acute{\beta}}+\mathbf{D}_{\alpha }\eta ^{\acute{\beta}}%
\mbox{ and }\mathbf{D}_{\alpha }(f\ \xi ^{\acute{\beta}})=f\ \ \mathbf{D}%
_{\alpha }\xi ^{\acute{\beta}}+\xi ^{\acute{\beta}}\ \mathbf{D}_{\alpha }f,
\end{equation*}%
for every $\xi ^{\acute{\beta}},\eta ^{\acute{\beta}}\in \mathcal{\gamma }^{%
\acute{\beta}}$ and $f$ being a scalar field on $\mathbf{V.}\mathcal{\ }$ It
is also required that one holds the Leibnitz rule
\begin{equation*}
(\mathbf{D}_{\alpha }\zeta _{\acute{\beta}})\eta ^{\acute{\beta}}=\mathbf{D}%
_{\alpha }(\zeta _{\acute{\beta}}\eta ^{\acute{\beta}})-\zeta _{\acute{\beta}%
}\ \mathbf{D}_{\alpha }\eta ^{\acute{\beta}}
\end{equation*}%
and that $\mathbf{D}_{\alpha }\,$ is a real operator, i. e. it commuters
with the operation of complex conjugation:
\begin{equation*}
\overline{\mathbf{D}_{\alpha }\ \psi _{\underline{\alpha }\underline{\beta }%
\underline{\gamma }...}}=\mathbf{D}_{\alpha }(\overline{\psi }_{\underline{%
\alpha }\underline{\beta }\underline{\gamma }...}).
\end{equation*}

Let us now analyze the question on uniqueness of action on d--spinors of an
operator $\mathbf{D}_{\alpha }$ satisfying some necessary conditions.
Denoting by $\mathbf{D}_{\alpha }^{(1)}$ and $\mathbf{D}_{\alpha }$ two such
d-covariant operators, we consider the map
\begin{equation}
(\mathbf{D}_{\alpha }^{(1)}-\mathbf{D}_{\alpha }):\ \mathcal{\gamma }^{%
\acute{\beta}}\rightarrow \mathcal{\gamma }_{\acute{\alpha}\grave{\alpha}}^{%
\acute{\beta}}.  \label{2.81}
\end{equation}%
Because the action on a scalar $f$ of both operators $\mathbf{D}_{\alpha
}^{(1)}$ and $\mathbf{D}_{\alpha }$ must be identical, i.e.
\begin{equation}
\mathbf{D}_{\alpha }^{(1)}f=\mathbf{D}_{\alpha }f,  \label{2.82}
\end{equation}%
the action (\ref{2.81}) on $f=\omega _{\acute{\beta}}\xi ^{\acute{\beta}}$
must be written as
\begin{equation*}
(\mathbf{D}_{\alpha }^{(1)}-\mathbf{D}_{\alpha })(\omega _{\acute{\beta}}\xi
^{\acute{\beta}})=0.
\end{equation*}%
We conclude that there is an element $\Theta _{\acute{\alpha}\grave{\alpha}%
\acute{\beta}}^{\quad \ \acute{\gamma}}\in \mathcal{\gamma }_{\acute{\alpha}%
\grave{\alpha}\acute{\beta}}^{\quad \ \acute{\gamma}}$ for which
\begin{equation}
\mathbf{D}_{\acute{\alpha}\grave{\alpha}}^{(1)}\ \xi ^{\acute{\gamma}}=%
\mathbf{D}_{\acute{\alpha}\grave{\alpha}}\xi ^{\acute{\gamma}}+\Theta _{%
\acute{\alpha}\grave{\alpha}\acute{\beta}}^{\quad \ \acute{\gamma}}\xi ^{%
\acute{\beta}}  \label{2.83}
\end{equation}%
and
\begin{equation*}
\mathbf{D}_{\acute{\alpha}\grave{\alpha}}^{(1)}\omega _{\acute{\beta}}=%
\mathbf{D}_{\acute{\alpha}\grave{\alpha}}\omega _{\acute{\beta}}-\Theta _{%
\acute{\alpha}\grave{\alpha}\acute{\beta}}^{\quad \ \acute{\gamma}}\omega _{%
\acute{\gamma}}~.
\end{equation*}%
The action of the operator (\ref{2.81}) on a d-vector $v^{\beta }=v^{\acute{%
\beta}\grave{\beta}}$ can be written by using formula (\ref{2.83}) for both
indices $\acute{\beta}$ and $\grave{\beta}$ :
\begin{eqnarray*}
(\mathbf{D}_{\alpha }^{(1)}-\mathbf{D}_{\alpha })v^{\acute{\beta}\grave{\beta%
}} &=&\Theta _{\alpha \acute{\gamma}}^{\quad \acute{\beta}}v^{\acute{\gamma}%
\grave{\beta}}+\Theta _{\alpha \grave{\gamma}}^{\quad \grave{\beta}}v^{%
\acute{\beta}\grave{\gamma}} \\
&=&(\Theta _{\alpha \acute{\gamma}}^{\quad \acute{\beta}}e_{\grave{\gamma}}^{%
\grave{\beta}}+\Theta _{\alpha \grave{\gamma}}^{\quad \grave{\beta}}e_{%
\acute{\gamma}}^{\acute{\beta}})v^{\acute{\gamma}\grave{\gamma}}=Q_{\ \alpha
\gamma }^{\beta }v^{\gamma },
\end{eqnarray*}%
where
\begin{equation}
Q_{\ \alpha \gamma }^{\beta }=Q_{~~\acute{\alpha}\grave{\alpha}~\acute{\gamma%
}\grave{\gamma}}^{\acute{\beta}\grave{\beta}}=\Theta _{\alpha \acute{\gamma}%
}^{\quad \acute{\beta}}e_{\grave{\gamma}}^{\grave{\beta}}+\Theta _{\alpha
\grave{\gamma}}^{\quad \grave{\beta}}e_{\acute{\gamma}}^{\acute{\beta}}.
\label{2.84}
\end{equation}%
The commutator $\mathbf{D}_{[\alpha }\mathbf{D}_{\beta ]}$ defines the
d--torsion. Applying operators $\mathbf{D}_{[\alpha }^{(1)}\mathbf{D}_{\beta
]}^{(1)}$ and $\mathbf{D}_{[\alpha }\mathbf{D}_{\beta ]}$ on $f=\omega _{%
\acute{\beta}}\xi ^{\acute{\beta}},$ we can write
\begin{equation*}
T_{\quad \alpha \beta }^{(1)\gamma }-T_{~\alpha \beta }^{\gamma }=Q_{~\beta
\alpha }^{\gamma }-Q_{~\alpha \beta }^{\gamma }
\end{equation*}%
with $Q_{~\alpha \beta }^{\gamma }$ from (\ref{2.84}).

The action of operator $\mathbf{D}_{\alpha }^{(1)}$ on d--spinor tensors
must be constructed by using formula (\ref{2.83}) for every upper indices
and formula (\ref{2.84}) for every lower indices.

\subsubsection{N--adapted Infeld - van der Waerden coefficients}

A d--spinor $\kappa ^{\acute{\alpha}}\in \mathcal{\gamma }$ $^{\acute{\alpha}%
} $ has the components $\kappa ^{\mathbf{\acute{\alpha}}}=\kappa ^{\acute{%
\alpha}}\mathbf{e}_{\acute{\alpha}}^{\mathbf{\acute{\alpha}}}=(\kappa ^{%
\mathbf{\acute{\imath}}},\kappa ^{\mathbf{\acute{a}}})$ defined with respect
to the N--adapted spinor basis (\ref{dsb}). Taking into account that
\begin{equation*}
\mathbf{e}_{\mathbf{\acute{\alpha}}}^{\ \acute{\alpha}}\ \mathbf{e}_{\mathbf{%
\grave{\beta}}}^{\ \grave{\beta}}\ \mathbf{D}_{\acute{\alpha}\grave{\beta}}=%
\mathbf{D}_{\mathbf{\acute{\alpha}\grave{\beta}}},
\end{equation*}%
we compute the components $\mathbf{D}_{\acute{\alpha}\grave{\beta}}$ $\kappa
^{\underline{\gamma }},$
\begin{eqnarray}
\mathbf{e}_{\mathbf{\acute{\alpha}}}^{\ \acute{\alpha}}\ \mathbf{e}_{\mathbf{%
\grave{\beta}}}^{\ \grave{\beta}}~\mathbf{e}_{\acute{\gamma}}^{\ \mathbf{%
\acute{\gamma}}}~\mathbf{D}_{\acute{\alpha}\grave{\beta}}\kappa ^{\acute{%
\gamma}} &=&\mathbf{e}_{\mathbf{\acute{\epsilon}}}^{\ \acute{\tau}}~\mathbf{e%
}_{\acute{\tau}}^{\ \mathbf{\acute{\gamma}}}~\mathbf{D}_{\mathbf{\acute{%
\alpha}}\underline{\mathbf{\grave{\beta}}}}\kappa ^{\mathbf{\acute{\epsilon}}%
}+\kappa ^{\mathbf{\acute{\epsilon}}}~\mathbf{e}_{\acute{\epsilon}}^{\
\mathbf{\acute{\gamma}}}~\mathbf{D}_{\mathbf{\acute{\alpha}\grave{\beta}}}%
\mathbf{e}_{\mathbf{\acute{\epsilon}}}^{\ \acute{\epsilon}}  \label{2.85} \\
&=&\mathbf{D}_{\mathbf{\acute{\alpha}\grave{\beta}}}\kappa ^{\mathbf{\acute{%
\gamma}}}+\kappa ^{\mathbf{\acute{\epsilon}}}\mathbf{\varpi }_{~\mathbf{%
\acute{\alpha}\grave{\beta}\acute{\epsilon}}}^{\mathbf{\acute{\gamma}}},
\notag
\end{eqnarray}%
where the coordinate components of the d--spinor connection are defined
\begin{equation}
\varpi _{~\mathbf{\acute{\alpha}\grave{\beta}\acute{\epsilon}}}^{\mathbf{%
\acute{\gamma}}}\doteq \mathbf{e}_{\acute{\tau}}^{\ \mathbf{\acute{\gamma}}}~%
\mathbf{D}_{\mathbf{\acute{\alpha}\grave{\beta}}}\mathbf{e}_{\mathbf{\acute{%
\epsilon}}}^{\ \acute{\tau}}.  \label{2.86}
\end{equation}%
We call the Infeld - van der Waerden d-symbols a set of objects $\varpi _{~%
\mathbf{\acute{\alpha}\grave{\beta}\acute{\epsilon}}}^{\mathbf{\acute{\gamma}%
}}$ paramet\-ri\-zed with respect to a coordinate d--spinor basis. Defining $%
\mathbf{D}_{\mathbf{\alpha }}=(\gamma _{\mathbf{\alpha }})^{\acute{\alpha}%
\grave{\beta}}~\mathbf{D}_{\acute{\alpha}\grave{\beta}},$ introducing
denotations $\varpi ^{\mathbf{\acute{\gamma}}}{}_{\mathbf{\alpha \acute{\tau}%
}}\doteq \varpi _{~\mathbf{\acute{\alpha}\grave{\beta}}\acute{\tau}}^{%
\mathbf{\acute{\gamma}}}~(\gamma _{\mathbf{\alpha }})^{\mathbf{\acute{\alpha}%
\grave{\beta}}}$ and using properties (\ref{2.85}), we write the relations
\begin{equation}
l_{\mathbf{\alpha }}^{\alpha }~\mathbf{e}_{\acute{\beta}}^{\ \mathbf{\acute{%
\beta}}}~\mathbf{D}_{\alpha }\kappa ^{\acute{\beta}}=\mathbf{D}_{\mathbf{%
\alpha }}\kappa ^{\mathbf{\acute{\beta}}}+\kappa ^{\mathbf{\acute{\delta}}%
}~\varpi _{~\mathbf{\alpha \acute{\delta}}}^{\mathbf{\acute{\beta}}}
\label{2.87}
\end{equation}%
and
\begin{equation}
l_{\mathbf{\alpha }}^{\alpha }~\mathbf{e}_{\mathbf{\acute{\beta}}}^{\ \acute{%
\beta}}~\mathbf{D}_{\alpha }~\mu _{\acute{\beta}}=\mathbf{D}_{\mathbf{\alpha
}}~\mu _{\mathbf{\acute{\beta}}}-\mu _{\mathbf{\acute{\delta}}}\varpi _{~%
\mathbf{\alpha \acute{\beta}}}^{\mathbf{\acute{\delta}}}  \label{2.88}
\end{equation}%
for d-covariant derivations $~\mathbf{D}_{\alpha }\kappa ^{\acute{\beta}}$
and $\mathbf{D}_{\alpha }~\mu _{\acute{\beta}}.$

We can consider expressions similar to (\ref{2.87}) and (\ref{2.88}) for
values having both types of d--spinor and d--tensor indices, for instance,
\begin{equation*}
l_{\mathbf{\alpha }}^{\alpha }~l_{\gamma }^{\mathbf{\gamma }}~\mathbf{e}_{%
\mathbf{\acute{\delta}}}^{\ \mathbf{\acute{\delta}}}~\mathbf{D}_{\alpha
}\theta _{\acute{\delta}}^{~\gamma }=\mathbf{D}_{\mathbf{\alpha }}\theta _{%
\mathbf{\acute{\delta}}}^{~\mathbf{\gamma }}-\theta _{\mathbf{\acute{\epsilon%
}}}^{~\mathbf{\gamma }}\ \varpi _{~\mathbf{\alpha \acute{\delta}}}^{\mathbf{%
\acute{\epsilon}}}+\theta _{\mathbf{\acute{\delta}}}^{~\mathbf{\tau }%
}~\Gamma _{\quad \mathbf{\alpha \tau }}^{~\mathbf{\gamma }}
\end{equation*}%
(we can prove this by a straightforward calculation of the derivation $%
\mathbf{D}_{\mathbf{\alpha }}(\theta _{\mathbf{\acute{\delta}}}^{~\mathbf{%
\tau }}$ $~\mathbf{e}_{\mathbf{\acute{\delta}}}^{\ \mathbf{\acute{\delta}}%
}~l_{\mathbf{\tau }}^{\gamma })).$

Now we shall consider some possible relations between components of
d--connec\-ti\-ons $\varpi _{~\mathbf{\alpha \acute{\delta}}}^{\mathbf{%
\acute{\epsilon}}}$ and $\Gamma _{\quad \mathbf{\alpha \tau }}^{~\mathbf{%
\gamma }}$ and derivations of $(\gamma _{\mathbf{\alpha }})^{\acute{\alpha}%
\grave{\beta}}.$ We can write
\begin{eqnarray*}
\Gamma _{~\mathbf{\beta \tau }}^{\mathbf{\alpha }} &=&l_{\alpha }^{\mathbf{%
\alpha }}\mathbf{D}_{\mathbf{\tau }}l_{\mathbf{\beta }}^{\alpha }=l_{\alpha
}^{\mathbf{\alpha }}\mathbf{D}_{\mathbf{\tau }}(\gamma _{\mathbf{\beta }})^{%
\acute{\epsilon}\grave{\tau}}=l_{\alpha }^{\mathbf{\alpha }}\mathbf{D}_{%
\mathbf{\tau }}((\gamma _{\mathbf{\beta }})^{\mathbf{\acute{\epsilon}\grave{%
\tau}}}\ \mathbf{e}_{\mathbf{\acute{\epsilon}}}^{\ \acute{\epsilon}}\
\mathbf{e}_{\mathbf{\grave{\tau}}}^{\ \grave{\tau}}) \\
&=&l_{\alpha }^{\mathbf{\alpha }}\mathbf{e}_{\mathbf{\acute{\alpha}}}^{\
\acute{\alpha}}\mathbf{e}_{\mathbf{\grave{\epsilon}}}^{\ \grave{\epsilon}}%
\mathbf{D}_{\mathbf{\tau }}(\gamma _{\mathbf{\beta }})^{\mathbf{\acute{\alpha%
}\grave{\epsilon}}}+l_{\alpha }^{\mathbf{\alpha }}(\gamma _{\mathbf{\beta }%
})^{\mathbf{\acute{\epsilon}\grave{\tau}}}(\mathbf{e}_{\mathbf{\grave{\tau}}%
}^{\ \grave{\tau}}\mathbf{D}_{\mathbf{\tau }}\mathbf{e}_{\mathbf{\acute{%
\epsilon}}}^{\ \acute{\epsilon}}+\mathbf{e}_{\mathbf{\acute{\epsilon}}}^{\
\acute{\epsilon}}\ \mathbf{D}_{\mathbf{\tau }}\mathbf{e}_{\mathbf{\grave{\tau%
}}}^{\ \grave{\tau}}) \\
&=&l_{\mathbf{\acute{\alpha}\grave{\epsilon}}}^{\mathbf{\alpha }}~\mathbf{D}%
_{\mathbf{\tau }}(\gamma _{\mathbf{\beta }})^{\mathbf{\acute{\alpha}\grave{%
\epsilon}}}+l_{\acute{\alpha}\grave{\epsilon}}^{\mathbf{\alpha }}\mathbf{e}_{%
\mathbf{\acute{\alpha}}}^{\ \acute{\alpha}}\mathbf{e}_{\mathbf{\grave{%
\epsilon}}}^{\ \grave{\epsilon}}(\gamma _{\mathbf{\beta }})^{\mathbf{\acute{%
\epsilon}\grave{\tau}}}(\mathbf{e}_{\mathbf{\grave{\tau}}}^{\ \grave{\tau}}%
\mathbf{D}_{\mathbf{\tau }}\mathbf{e}_{\mathbf{\acute{\epsilon}}}^{\ \acute{%
\epsilon}}+\mathbf{e}_{\mathbf{\acute{\epsilon}}}^{\ \acute{\epsilon}}\
\mathbf{D}_{\mathbf{\tau }}\mathbf{e}_{\mathbf{\grave{\tau}}}^{\ \grave{\tau}%
}),
\end{eqnarray*}%
where $l_{\alpha }^{\mathbf{\alpha }}=(\gamma _{\acute{\alpha}\grave{\alpha}%
})^{\mathbf{\alpha }}$ , from which one follows
\begin{equation*}
(\gamma _{\mathbf{\alpha }})^{\mathbf{\acute{\mu}}\grave{\nu}}(\gamma ^{%
\mathbf{\beta }})_{\mathbf{\acute{\alpha}\grave{\beta}}}\Gamma _{~\mathbf{%
\gamma \beta }}^{\mathbf{\alpha }}=(\gamma ^{\mathbf{\beta }})_{\mathbf{%
\acute{\alpha}\grave{\beta}}}\mathbf{D}_{\mathbf{\gamma }}(\gamma _{\mathbf{%
\beta }})^{\mathbf{\acute{\mu}}\grave{\nu}}+\mathbf{e}_{\mathbf{\grave{\beta}%
}}^{\ \grave{\nu}}\varpi _{~\mathbf{\gamma \acute{\alpha}}}^{\mathbf{\acute{%
\mu}}}+\mathbf{e}_{\mathbf{\acute{\alpha}}}^{\ \mathbf{\acute{\mu}}}\varpi
_{~\mathbf{\gamma \grave{\beta}}}^{\grave{\nu}}.
\end{equation*}%
Contracting the last expression on $\grave{\nu}$ and $\grave{\beta}$ and
using an orthonormalized d--spinor basis when $\varpi _{~\mathbf{\gamma
\acute{\beta}}}^{\mathbf{\acute{\beta}}}=0$ (a consequence from (\ref{2.86}%
)), we have
\begin{equation}
\varpi _{~\mathbf{\gamma }\acute{\alpha}}^{\mathbf{\acute{\mu}}}=\frac{1}{%
\tilde{k}_{(n)}+\tilde{k}_{(m)}}(\Gamma _{\quad \mathbf{\gamma ~\acute{\alpha%
}\grave{\beta}}}^{\mathbf{\acute{\mu}\grave{\beta}}}-(\gamma ^{\mathbf{\beta
}})_{\acute{\alpha}\mathbf{\grave{\beta}}}\mathbf{D}_{\mathbf{\gamma }%
}(\gamma _{\mathbf{\beta }})^{\mathbf{\acute{\mu}\grave{\beta}}}),
\label{2.89}
\end{equation}%
where
\begin{equation}
\Gamma _{\quad \mathbf{\gamma ~\acute{\alpha}\grave{\beta}}}^{\mathbf{\acute{%
\mu}\grave{\beta}}}=(\gamma _{\mathbf{\alpha }})^{\mathbf{\acute{\mu}\grave{%
\beta}}}(\gamma ^{\mathbf{\beta }})_{\acute{\alpha}\mathbf{\grave{\beta}}%
}\Gamma _{~\mathbf{\gamma \beta }}^{\mathbf{\alpha }}.  \label{2.90}
\end{equation}%
The d--spinor connection (\ref{2.90}) can be defined by various type of of
d--connecti\-ons, inclusively, by the canonical one, see
 \cite{vncsym}. Such formulas can be applied on Clifford algebroid $%
\mathcal{C}(E)\doteqdot (\mathbb{C}l(E),\ ^{s}\left[ \cdot ,\cdot \right] ,\
^{s}\rho )$ or on a Clifford N--aglebroid $\mathcal{C}(v\mathbf{V})\doteqdot
(\mathbb{C}l(v\mathbf{V}),\ ^{s}\left[ \cdot ,\cdot \right] ,\ ^{s}\rho ).$
We have to change the v--derivatives into anchored
ones, $\partial /\partial y^{a}\rightarrow \rho _{a}^{i}\partial /\partial
x^{a},$ or in N--adapted form, $e_{a}\rightarrow {}\widehat{\mathbf{\rho }}%
_{a}^{j}e_{j},$ and put the results in formulas (\ref{2.89}) \ and (\ref%
{2.90}). $\ $In result, one defines a canonical covariant spinor
differential calculus, adapted to the N--connection structure, acting on the
set of sections $Sec(\mathbf{E})$ or $Sec(v\mathbf{V}).$

\subsubsection{ D--spinors of curvature and torsion on N--anholonomic
manifolds}

The d-tensor indices of the commutator $\Delta _{\alpha \beta }$ can be
transformed into d--spinor ones:
\begin{equation}
\square _{\acute{\alpha}\grave{\beta}}=(\gamma ^{\alpha \beta })_{\acute{%
\alpha}\grave{\beta}}\Delta _{\alpha \beta }=(\square _{\acute{\imath}\grave{%
\imath}},\square _{\acute{a}\grave{a}}),  \label{2.91}
\end{equation}%
with h- and v-components,
\begin{equation*}
\square _{\acute{\imath}\grave{\imath}}=(\gamma ^{\alpha \beta })_{\acute{%
\imath}\grave{\imath}}\Delta _{\alpha \beta }\mbox{ and }\square _{\acute{a}%
\grave{a}}=(\gamma ^{\alpha \beta })_{\acute{a}\grave{a}}\Delta _{\alpha
\beta },
\end{equation*}%
being symmetric or antisymmetric in dependence of corresponding values of
dimensions $n\,$ and $m.$
Considering the actions of operator (\ref{2.91}) on d--spinors $%
\pi ^{\acute{\gamma}}$ and $\mu _{\acute{\gamma}}$ we introduce the
d--spinor curvature $X_{\ \acute{\tau}\acute{\alpha}\grave{\beta}}^{\acute{%
\gamma}\quad }$ satisfying the equations
\begin{equation}
\square _{\acute{\alpha}\grave{\beta}}\ \pi ^{\acute{\gamma}}=X_{\ \acute{%
\tau}\acute{\alpha}\grave{\beta}}^{\acute{\gamma}\quad }\pi ^{\acute{\tau}}
\label{2.92}
\end{equation}%
and
\begin{equation*}
\square _{\acute{\alpha}\grave{\beta}}\mu _{\acute{\gamma}}=X_{\ \acute{%
\gamma}\acute{\alpha}\grave{\beta}}^{\acute{\tau}\quad }\mu _{\acute{\tau}}.
\end{equation*}%
The gravitational d--spinor $\Psi _{\acute{\tau}\acute{\gamma}\acute{\alpha}%
\grave{\beta}}$ is defined by a corresponding symmetrization of d--spinor
indices:
\begin{equation}
\Psi _{\acute{\tau}\acute{\gamma}\acute{\alpha}\grave{\beta}}=X_{\ (\acute{%
\tau}\acute{\gamma}\acute{\alpha})\grave{\beta}}.  \label{2.93}
\end{equation}%
We note that d--spinor tensors $X_{\ \acute{\gamma}\acute{\alpha}\grave{\beta%
}}^{\acute{\tau}\quad }$ and $\Psi _{\acute{\tau}\acute{\gamma}\acute{\alpha}%
\grave{\beta}}\,$ are transformed into similar 2--spinor objects if the
N--connection vanishes and the spinor constuctions are defined in global
form on $\mathbf{V}$ \cite{penr1,penr2}.

Putting $\mathbf{e}_{\acute{\gamma}}^{\quad \mathbf{\acute{\gamma}}}$
instead of $\mu _{\acute{\gamma}}$ in (\ref{2.92}) and using (\ref{2.93}),
we can express respectively the curvature and gravitational d--spinors as
\begin{equation*}
X_{\acute{\gamma}\acute{\delta}\acute{\alpha}\grave{\beta}}=\mathbf{e}_{%
\acute{\gamma}\mathbf{\acute{\tau}}}\square _{\acute{\alpha}\grave{\beta}}%
\mathbf{e}_{\acute{\delta}}^{\ \mathbf{\acute{\tau}}}
\end{equation*}%
and
\begin{equation*}
\Psi _{\acute{\delta}\acute{\gamma}\acute{\alpha}\grave{\beta}}=\mathbf{e}_{%
\acute{\delta}(\mathbf{\acute{\tau}}}\square _{\acute{\alpha}|\grave{\beta}|}%
\mathbf{e}_{\acute{\gamma})}^{\ \mathbf{\acute{\tau}}}
\end{equation*}%
where we omit symmetrization on $\grave{\beta}.$

The d--spinor torsion $T_{\quad \acute{\alpha}\grave{\beta}}^{\acute{\gamma}%
\grave{\gamma}}$ is defined by using the d--spinor commutator (\ref{2.91})
and equations
\begin{equation*}
\square _{\acute{\alpha}\grave{\beta}}f=T_{\quad \acute{\alpha}\grave{\beta}%
}^{\acute{\gamma}\grave{\gamma}}\bigtriangledown _{\acute{\gamma}\grave{%
\gamma}}f.
\end{equation*}

The d--spinor components $R_{\quad \acute{\gamma}\grave{\gamma}\acute{\alpha}%
\grave{\beta}}^{\acute{\delta}\grave{\delta}\qquad }$ of the curvature
d-tensor $R_{\ \gamma \alpha \beta }^{\delta }$ can be computed by using the
relations (\ref{2.90}), (\ref{2.91}) and (\ref{2.93}) and the equations
\begin{equation}
(\square _{\acute{\alpha}\grave{\beta}}-T_{\quad \acute{\alpha}\grave{\beta}%
}^{\acute{\gamma}\grave{\gamma}}\bigtriangledown _{\acute{\gamma}\grave{%
\gamma}})V^{\acute{\delta}\grave{\delta}}=R_{\quad \acute{\gamma}\grave{%
\gamma}\acute{\alpha}\grave{\beta}}^{\acute{\delta}\grave{\delta}}V^{\acute{%
\gamma}\grave{\gamma}},  \label{2.94}
\end{equation}%
here d-vector $V^{\acute{\gamma}\grave{\gamma}}$ is considered as a product
of d--spinors, i. e. $V^{\acute{\gamma}\grave{\gamma}}=\nu ^{\acute{\gamma}%
}\mu ^{\grave{\gamma}}.$ We find
\begin{equation}
R_{\quad \acute{\gamma}\grave{\gamma}\acute{\alpha}\grave{\beta}}^{\acute{%
\delta}\grave{\delta}}=\left( X_{\quad \acute{\gamma}\acute{\alpha}\grave{%
\beta}}^{\acute{\delta}}+T_{\quad \acute{\alpha}\grave{\beta}}^{\acute{\tau}%
\grave{\tau}}\ \varpi _{\ \acute{\tau}\grave{\tau}\acute{\gamma}}^{\acute{%
\delta}}\right) \mathbf{e}_{\grave{\gamma}}^{\ \grave{\delta}}+\left(
X_{\quad \grave{\gamma}\acute{\alpha}\grave{\beta}}^{\grave{\delta}%
}+T_{\quad \acute{\alpha}\grave{\beta}}^{\acute{\tau}\grave{\tau}}\ \varpi
_{\ \acute{\tau}\grave{\tau}\grave{\gamma}}^{\grave{\delta}}\right) \mathbf{e%
}_{\acute{\gamma}}^{\ \acute{\delta}}.  \notag
\end{equation}%
It is convenient to use this d--spinor expression for the curvature
d--tensor in order to get the d--spinor components of the
Ricci d--tensor,
\begin{equation}
R_{\acute{\gamma}\grave{\gamma}\acute{\alpha}\grave{\beta}}=R_{\quad \acute{%
\gamma}\grave{\gamma}\acute{\alpha}\grave{\beta}\acute{\delta}\grave{\delta}%
}^{\acute{\delta}\grave{\delta}}=X_{\ \ \acute{\gamma}\acute{\alpha}\grave{%
\beta}\acute{\delta}\grave{\gamma}}^{\acute{\delta}}+T_{\quad \acute{\alpha}%
\grave{\beta}\acute{\delta}\grave{\gamma}}^{\acute{\tau}\grave{\tau}}\
\varpi _{\ \acute{\tau}\grave{\tau}\acute{\gamma}}^{\acute{\delta}}+X_{\
\grave{\gamma}\acute{\alpha}\grave{\beta}\acute{\gamma}\grave{\delta}}^{%
\grave{\delta}}+T_{\quad \acute{\alpha}\grave{\beta}\acute{\gamma}\grave{%
\delta}}^{\acute{\tau}\grave{\tau}}\ \varpi _{\ \acute{\tau}\grave{\tau}%
\grave{\gamma}}^{\grave{\delta}}  \label{sricci}
\end{equation}%
and this d--spinor decomposition of the scalar curvature $\overleftarrow{R}%
=R_{\quad \acute{\alpha}\grave{\beta}}^{\acute{\alpha}\grave{\beta}},$
\begin{equation*}
\overleftarrow{R}= R_{\quad ~\acute{\alpha}\grave{\beta}\acute{\delta}\grave{%
\delta}}^{\acute{\delta}\grave{\delta}\acute{\alpha}\grave{\beta}}=X_{\ \
\acute{\alpha}\grave{\beta}\acute{\delta}}^{\acute{\delta}\acute{\alpha}%
\quad ~\grave{\beta}}+T_{\quad \acute{\alpha}\grave{\beta}\acute{\delta}}^{%
\acute{\tau}\grave{\tau}\quad ~\grave{\beta}}\ \varpi _{\ \acute{\tau}\grave{%
\tau}}^{\acute{\delta}~\ \acute{\alpha}}+X_{\ ~\acute{\alpha}\grave{\beta}\
\grave{\delta}}^{\grave{\delta}\grave{\beta}~\ \acute{\alpha}}+T_{\quad
\acute{\alpha}\grave{\beta}\ \grave{\delta}}^{\acute{\tau}\grave{\tau}~\
\acute{\alpha}}\ \varpi _{\ \acute{\tau}\grave{\tau}}^{\grave{\delta}~~%
\grave{\beta}}.
\end{equation*}%
Finally, we write down the d--spinor components of the Einstein d--tensor $%
\mathbf{G}_{\gamma \beta },$
\begin{eqnarray}
&& \mathbf{G}_{\acute{\gamma}\grave{\gamma}\acute{\beta}\grave{\beta}} =
X_{\ \ \acute{\gamma}\acute{\beta}\grave{\beta}\acute{\delta}\grave{\gamma}%
}^{\acute{\delta}}+T_{\quad \acute{\beta}\grave{\beta}\acute{\delta}\grave{%
\gamma}}^{\acute{\tau}\grave{\tau}}\ \varpi _{\ \acute{\tau}\grave{\tau}%
\acute{\gamma}}^{\acute{\delta}}+X_{\ \grave{\gamma}\acute{\beta}\grave{\beta%
}\acute{\gamma}\grave{\delta}}^{\grave{\delta}}+T_{\quad \acute{\beta}\grave{%
\beta}\acute{\gamma}\grave{\delta}}^{\acute{\tau}\grave{\tau}}\ \varpi _{\
\acute{\tau}\grave{\tau}\grave{\gamma}}^{\grave{\delta}}  \label{seinst} \\
&&-\frac{1}{2}\epsilon _{\acute{\gamma}\grave{\beta}}\epsilon _{\acute{\beta}%
\grave{\gamma}}\left[ X_{\ \ \acute{\alpha}\grave{\beta}\acute{\delta}}^{%
\acute{\delta}\acute{\alpha}\quad ~\grave{\beta}}+T_{\quad \acute{\alpha}%
\grave{\beta}\acute{\delta}}^{\acute{\tau}\grave{\tau}\quad ~\grave{\beta}}\
\varpi _{\ \acute{\tau}\grave{\tau}}^{\acute{\delta}~\ \acute{\alpha}}+X_{\ ~%
\acute{\alpha}\grave{\beta}\ \grave{\delta}}^{\grave{\delta}\grave{\beta}~\
\acute{\alpha}}+T_{\quad \acute{\alpha}\grave{\beta}\ \grave{\delta}}^{%
\acute{\tau}\grave{\tau}~\ \acute{\alpha}}\ \varpi _{\ \acute{\tau}\grave{%
\tau}}^{\grave{\delta}~~\grave{\beta}}\right] .  \notag
\end{eqnarray}

It should be noted that further reductions of (\ref{sricci}) and (\ref%
{seinst}) depend on dimensions $n$ and $m$ of the, respectively, h-- and
v--subspaces, and that the symmetry properties are defined by the $\epsilon $%
--objects.  On Clifford N--algebroids,
such formulas have to be considered for anchored v--derivatives (\ref{anch1d}%
) and (\ref{anch}) (for d--spinor considerations, we have to apply spinor
anchors (\ref{anchs}) and (\ref{anchps})), for instance, in the case of
canonical d--connections  and their spinor variants (\ref{2.89}).

\section{Field Equations and Lie Algebro\-ids}

Lie algebroid structures can be modelled as spacetime geometries with
generalized symmetries (defined by anchors and Lie algebra commutators and
nontrivial N--connection structure) \cite{valgexsol}. It is possible to
extend the constructions on Clifford N--algebroids by introducing spinor
variables.\thinspace\ In this section we shall analyze the basic field
equations for gravitational and matter field interactions modelled on
N--anholonomic manifolds and Clifford N--algebroids.

\subsection{The Dirac operator on N--anholonomic spaces}

\label{sdonst} The aim of this section is to elucidate the possibility of
definition of Dirac operators for general N--anholonomic manifolds. It
should be noted that such geometric constructions depend on the type of
linear connections which are used for the complete definition of the Dirac
operator. They are metric compatible and N--adapted if the canonical
d--connection is used (we can similarly use any its deformation resulting in
a metric compatible d--connection).

\subsubsection{Noholonomic vielbeins and spin d--connections}

For a local dual coordinate basis $e^{\underline{i}}\doteq dx^{\underline{i}%
} $ on a manifold $M,\ dim\ M=n,$\ we may respectively introduce certain
classes of orthonormalized vielbeins and the N--adapted vielbeins (depending
both on the base coordinates $x\doteq \ x^{i}$ and some ''fiber''
coordinates $y\doteq y^{a}$)
\begin{equation}
e^{\hat{\imath}}\doteq e_{\ \underline{i}}^{\hat{\imath}}(x,y)\ e^{%
\underline{i}}\mbox{
and }e^{i}\doteq e_{\ \underline{i}}^{i}(x,y)\ e^{\underline{i}},
\label{hatbvb}
\end{equation}%
where
\begin{equation*}
g^{\underline{i}\underline{j}}(x,y)\ e_{\ \underline{i}}^{\hat{\imath}%
}(x,y)e_{\ \underline{j}}^{\hat{\jmath}}(x,y)=\delta ^{\hat{\imath}\hat{%
\jmath}}\mbox{ and }g^{\underline{i}\underline{j}}(x,y)\ e_{\ \underline{i}%
}^{i}(x,y)e_{\ \underline{j}}^{j}(x,y)=g^{ij}(x,y).
\end{equation*}%
We define the the algebra of Dirac's gamma matrices (in brief, h--gamma
matrices defined by self--adjoints matrices $M_{k}(\mathbb{C})$ where $%
k=2^{n/2}$ is the dimension of the irreducible representation of $\mathbb{C}%
l(M)$ for even dimensions, or of $\mathbb{C}l(M)^{+}$ for odd dimensions)
from the relation
\begin{equation}
\gamma ^{\hat{\imath}}\gamma ^{\hat{\jmath}}+\gamma ^{\hat{\jmath}}\gamma ^{%
\hat{\imath}}=2\delta ^{\hat{\imath}\hat{\jmath}}\ \mathbb{I}.
\label{grelflat}
\end{equation}%
We can consider the action of $dx^{i}\in \mathbb{C}l(M)$ on a spinor $\psi
\in S$ via representations
\begin{equation}
\ ^{-}c(dx^{\hat{\imath}})\doteq \gamma ^{\hat{\imath}}\mbox{ and }\
^{-}c(dx^{i})\psi \doteq \gamma ^{i}\psi \equiv e_{\ \hat{\imath}}^{i}\
\gamma ^{\hat{\imath}}\psi .  \label{gamfibb}
\end{equation}

For any type of spaces $T_{x}M,TM$ or $\mathbf{V}$ possessing a local (in
any point) or global fibered structure and enabled with a N--connection
structure, we can introduce similar definitions of the gamma matrices
following algebraic relations and metric structures on fiber subspaces,
\begin{equation}
e^{\hat{a}}\doteq e_{\ \underline{a}}^{\hat{a}}(x,y)\ e^{\underline{a}}%
\mbox{
and }e^{a}\doteq e_{\ \underline{a}}^{a}(x,y)\ e^{\underline{a}},
\label{hatbvf}
\end{equation}%
where
\begin{equation*}
g^{\underline{a}\underline{b}}(x,y)\ e_{\ \underline{a}}^{\hat{a}}(x,y)e_{\
\underline{b}}^{\hat{b}}(x,y)=\delta ^{\hat{a}\hat{b}}\mbox{ and }g^{%
\underline{a}\underline{b}}(x,y)\ e_{\ \underline{a}}^{a}(x,y)e_{\
\underline{b}}^{b}(x,y)=h^{ab}(x,y).
\end{equation*}%
Similarly, we define the algebra of Dirac's matrices related to typical
fibers (in brief, v--gamma matrices described by self--adjoints matrices $%
M_{k}^{\prime }(\mathbb{C})$ where $k^{\prime }=2^{m/2}$ is the dimension of
the irreducible representation of $\mathbb{C}l(F)$ for even dimensions, or
of $\mathbb{C}l(F)^{+}$ for odd dimensions, of the typical fiber $F$) from
the relation
\begin{equation}
\gamma ^{\hat{a}}\gamma ^{\hat{b}}+\gamma ^{\hat{b}}\gamma ^{\hat{a}%
}=2\delta ^{\hat{a}\hat{b}}\ \mathbb{I}.  \label{grelflatf}
\end{equation}%
The formulas (\ref{grelflat}) and (\ref{grelflatf}) are respectively the h--
and v--components of the relation (\ref{2.78}) (with redefined the
coefficients which is more convenient for further constructions). \ The
action of $dy^{a}\in \mathbb{C}l(F)$ on a spinor $\ ^{\star }\psi \in \
^{\star }S$ is considered via representations
\begin{equation}
\ ^{\star }c(dy^{\hat{a}})\doteq \gamma ^{\hat{a}}\mbox{ and }\ ^{\star
}c(dy^{a})\ ^{\star }\psi \doteq \gamma ^{a}\ ^{\star }\psi \equiv e_{\
\hat{a}}^{a}\ \gamma ^{\hat{a}}\ ^{\star }\psi .  \label{gamfibf}
\end{equation}%
We note that additionally to formulas (\ref{gamfibb}) and (\ref{gamfibf}) we
may write respectively
\begin{equation*}
c(dx^{\underline{i}})\psi \doteq \gamma ^{\underline{i}}\psi \equiv e_{\
\hat{\imath}}^{\underline{i}}\ \gamma ^{\hat{\imath}}\psi \mbox{ and }c(dy^{%
\underline{a}})\ ^{\star }\psi \doteq \gamma ^{\underline{a}}\ ^{\star }\psi
\equiv e_{\ \hat{a}}^{\underline{a}}\ \gamma ^{\hat{a}}\ ^{\star }\psi
\end{equation*}%
but such operators are not adapted to the N--connection structure.

A more general gamma matrix calculus with distinguished gamma matrices (in
brief, d--gamma matrices) can be elaborated for any N--anholonomic manifold $%
\mathbf{V}$ provided with d--metric structure $\mathbf{g}=[g,^{\star }g]$
and for d--spinors $\breve{\psi}\doteq (\psi ,\ ^{\star }\psi )\in \mathbf{S}%
\doteq (S,\ ^{\star }S).$ Firstly, we should write in a unified form,
related to a d--metric (\ref{metr}), the formulas (\ref{hatbvb}) and (\ref%
{hatbvf}),
\begin{equation}
e^{\hat{\alpha}}\doteq e_{\ \underline{a}}^{\hat{\alpha}}(u)\ e^{\underline{%
\alpha }}\mbox{ and }e^{\alpha }\doteq e_{\ \underline{\alpha }}^{\alpha
}(u)\ e^{\underline{\alpha }},  \label{hatbvd}
\end{equation}%
where
\begin{equation*}
g^{\underline{\alpha }\underline{\beta }}(u)\ e_{\ \underline{\alpha }}^{%
\hat{\alpha}}(u)e_{\ \underline{\beta }}^{\hat{\beta}}(u)=\delta ^{\hat{%
\alpha}\hat{\beta}}\mbox{ and }g^{\underline{\alpha }\underline{\beta }}(u)\
e_{\ \underline{\alpha }}^{\alpha }(u)e_{\ \underline{\beta }}^{\beta
}(u)=g^{\alpha \beta }(u).
\end{equation*}%
The second step, is to consider gamma d--matrix relations (unifying (\ref%
{grelflat}) and (\ref{grelflatf}))
\begin{equation}
\gamma ^{\hat{\alpha}}\gamma ^{\hat{\beta}}+\gamma ^{\hat{\beta}}\gamma ^{%
\hat{\alpha}}=2\delta ^{\hat{\alpha}\hat{\beta}}\ \mathbb{I},
\label{grelflatd}
\end{equation}%
with the action of $du^{\alpha }\in \mathbb{C}l(\mathbf{V})$ on a d--spinor $%
\breve{\psi}\in \ \mathbf{S}$ resulting in distinguished irreducible
representations (unifying (\ref{gamfibb}) and (\ref{gamfibf}))
\begin{equation}
\mathbf{c}(du^{\hat{\alpha}})\doteq \gamma ^{\hat{\alpha}}\mbox{
and }\mathbf{c}=(du^{\alpha })\ \breve{\psi}\doteq \gamma ^{\alpha }\ \breve{%
\psi}\equiv e_{\ \hat{\alpha}}^{\alpha }\ \gamma ^{\hat{\alpha}}\ \breve{\psi%
}  \label{gamfibd}
\end{equation}%
which allows to write
\begin{equation}
\gamma ^{\alpha }(u)\gamma ^{\beta }(u)+\gamma ^{\beta }(u)\gamma ^{\alpha
}(u)=2g^{\alpha \beta }(u)\ \mathbb{I}.  \label{grelnam}
\end{equation}%
In the canonical representation, we can write in irreducible form $\breve{%
\gamma}\doteq \gamma \oplus \ ^{\star }\gamma $ and $\breve{\psi}\doteq \psi
\oplus \ ^{\star }\psi ,$ for instance, by using block type of h-- and
v--matrices, or, writing alternatively as couples of gamma and/or h-- and
v--spinor objects written in N--adapted form,
\begin{equation}
\gamma ^{\alpha }\doteq (\gamma ^{i},\gamma ^{a})\mbox{ and }\breve{\psi}%
\doteq (\psi ,\ ^{\star }\psi ).  \label{crgs}
\end{equation}%
The decomposition (\ref{grelnam}) holds with respect to a N--adapted
vielbein (\ref{dder}). We also note that for a spinor calculus, the indices
of spinor objects should be treated as abstract spinorial ones possessing
certain reducible, or irreducible, properties depending on the space
dimension. For simplicity, we shall consider that spinors like $\breve{\psi}%
,\psi ,\ ^{\star }\psi $ and all type of gamma objects can be enabled with
corresponding spinor indices running certain values which are different from
the usual coordinate space indices.

The spin connection $\nabla ^{S}$ for the Riemannian manifolds is induced by
the Levi--Civita connection $\ ^{\nabla }\Gamma ,$
\begin{equation}
\nabla ^{S}\doteq d-\frac{1}{4}\ ^{\nabla }\Gamma _{\ jk}^{i}\gamma
_{i}\gamma ^{j}\ dx^{k}.  \label{sclcc}
\end{equation}%
On N--anholonomic spaces, it is possible to define spin connections which
are N--adapted by replacing the Levi--Civita connection by any d--connection.

The canonical spin d--connection is defined by the canonical d--connecti\-on,
\begin{equation}
\widehat{\nabla }^{\mathbf{S}}\doteq \delta -\frac{1}{4}\ \widehat{\mathbf{%
\Gamma }}_{\ \beta \mu }^{\alpha }\gamma _{\alpha }\gamma ^{\beta }\delta
u^{\mu },  \label{csdc}
\end{equation}%
where the absolute differential $\delta $ acts in N--adapted form resulting
in 1--forms decomposed with respect to N--elongated differentials $\delta
u^{\mu }=(dx^{i},\delta y^{a})$ (\ref{ddif}).

We note that the canonical spin d--connection $\widehat{\nabla }^{\mathbf{S}%
} $ is metric compatible and contains nontrivial d--torsion coefficients
induced by the N--anholonomy relations. It is possible to introduce more
general spin d--connections ${\mathbf{D}}^{\mathbf{S}}$ by using the same
formula (\ref{csdc}) but for arbitrary metric compatible d--connection ${%
\mathbf{\Gamma }}_{\ \beta \mu }^{\alpha }.$ For the spaces provided with
generic off--diagonal metric structure (\ref{metr}) on a
N--anho\-lo\-nom\-ic manifold, there is a canonical spin d--connection (\ref%
{csdc}) induced by the off--diagonal metric coefficients with nontrivial $%
N_{\ i}^{a}$ and associated nonholonomic frames in gravity theories.

In a particular case of N--anholonomic manifolds of even dimensions, we can
define, for instance, the canonical spin d--connections for a local
modelling of a tangent bundle space with the canonical d--connection $%
\widehat{\mathbf{\Gamma }}_{\ \alpha \beta }^{\gamma }=(\widehat{L}_{jk}^{i},%
\widehat{B}_{jk}^{i}).$
The N--connection structure $N_{\ i}^{j}$ states a global h-- and
v--splitting of the spin d--connection operators, for instance,
\begin{equation}
\widehat{\nabla }\doteq \delta -\frac{1}{4}\ {\widehat{L}}_{\ jk}^{i}\gamma
_{i}\gamma ^{j}dx^{k}-\frac{1}{4}\ {\widehat{B}}_{\ bc}^{a}\gamma _{a}\gamma
^{b}\delta y^{c}.  \label{cslc}
\end{equation}%
So, any spin d--connection is a d--operator with conventional splitting of
action like ${\nabla }^{(\mathbf{S})}\equiv ({\ ^{-}{\nabla }}^{(\mathbf{S}%
)},{\ ^{\star }{\nabla }}^{(\mathbf{S})}),$ or ${\nabla }\equiv ({\ ^{-}{%
\nabla }},{\ ^{\star }{\nabla }}).$ For instance, for $\widehat{\nabla }%
\equiv ({\ ^{-}\widehat{\nabla }},{\ ^{\star }\widehat{\nabla }}),$ the
operators $\ ^{-}\widehat{\nabla }$ and $\ ^{\star }\widehat{\nabla }$ act
respectively on a h--spinor $\psi $ as
\begin{equation}
{\ ^{-}\widehat{\nabla }}\psi \doteq dx^{i}\ \frac{\delta \psi }{\partial
x^{i}}-dx^{k}\frac{1}{4}\ {\widehat{L}}_{\ jk}^{i}\gamma _{i}\gamma ^{j}\
\psi  \label{hdslop}
\end{equation}%
and
\begin{equation*}
{\ ^{\star }\widehat{\nabla }}\psi \doteq \delta y^{a}\ \frac{\partial \psi
}{\partial y^{a}}-\delta y^{c}\ \frac{1}{4}\ {\widehat{B}}_{\ bc}^{a}\gamma
_{a}\gamma ^{b}\ \psi
\end{equation*}%
being defined by the canonical d--connection, which (in its
turn) is completely defined by $N_{\ i}^{j}(x,y)$ and $g_{ij}(x,y).$

The operators (\ref{hdslop}) can be adapted to the Lie algebroid structure
by anchoring the partial v--derivatives. For instance,
\begin{eqnarray*}
\frac{\delta \psi }{\partial x^{i}}(x^{k},y^{b}(x^{j})) &=&\frac{\partial
\psi }{\partial x^{i}}-N_{i}^{a}\frac{\partial \psi }{\partial y^{a}}=\left(
\frac{\partial \psi }{\partial x^{i}}-N_{i}^{a}\rho _{a}^{k}(x^{j})\frac{%
\partial \psi }{\partial x^{k}}\right) \\
&=&\left( \frac{\partial \psi }{\partial x^{i}}-\ ^{\rho }N_{i}^{k}\frac{%
\partial \psi }{\partial x^{k}}\right) (x^{k},y^{b}(x^{j}))
\end{eqnarray*}%
where the anchor $\rho _{a}^{k}$ (\ref{anch}) induce a N--connection $^{\rho
}N_{i}^{k}\doteqdot N_{i}^{a}\rho _{a}^{k}.$ We can also perform a
N--adapted Clifford algebroid calculus by using the ''boldface'' algebroid $%
{}\widehat{\mathbf{\rho }}_{a}^{j}$ (\ref{anch1d}) with explicit dependence
on variables $y^{b},$%
\begin{eqnarray*}
\frac{\delta \psi }{\partial x^{i}}(x^{k},y^{b}) &=&\frac{\partial \psi }{%
\partial x^{i}}-N_{i}^{a}e_{a}\psi =\left( \frac{\partial \psi }{\partial
x^{i}}-N_{i}^{a}{}\widehat{\mathbf{\rho }}_{a}^{k}\frac{\partial \psi }{%
\partial x^{k}}\right) \\
&=&\left( \frac{\partial \psi }{\partial x^{i}}-\ ^{\rho }\hat{N}_{i}^{k}%
\frac{\partial \psi }{\partial x^{k}}\right)
\end{eqnarray*}%
for $^{\rho }\hat{N}_{i}^{k}=N_{i}^{a}{}\widehat{\mathbf{\rho }}_{a}^{k}.$
Such anchoring of partial/N--elongated derivatives has to be considered for
the canonical d--connection $\ {\widehat{L}}_{\ jk}^{i}$ and $\ {\widehat{B}}%
_{\ bc}^{a}.$

\subsubsection{Dirac d--operators}

We consider a vector bundle $\mathbf{E}$ on an N--anholonomic manifold $M$
(with two compatible N--connections defined as h-- and v--splittings of $T%
\mathbf{E}$ and $TM$)). A d--connection
\begin{equation*}
\mathcal{D}:\ Sec^{\infty }(\mathbf{E})\rightarrow Sec^{\infty }(\mathbf{E}%
)\otimes \Omega ^{1}(M)
\end{equation*}%
preserves by parallelism splitting of the tangent total and base spaces and
satisfy the Leibniz condition
\begin{equation*}
\mathcal{D}(f\sigma )=f(\mathcal{D}\sigma )+\delta f\otimes \sigma
\end{equation*}%
for any $f\in C^{\infty }(M),$ and $\sigma \in Sec^{\infty }(\mathbf{E})$
and $\delta $ defining an N--adapted exterior calculus by using N--elongated
operators (\ref{dder}) and (\ref{ddif}) which emphasize d--forms instead of
usual forms on $M,$ with the coefficients taking values in $\mathbf{E}.$

The metricity and Leibniz conditions for $\mathcal{D}$ are written
respectively
\begin{equation}
\mathbf{g}(\mathcal{D}\mathbf{X},\mathbf{Y})+\mathbf{g}(\mathbf{X},\mathcal{D%
}\mathbf{Y})=\delta \lbrack \mathbf{g}(\mathbf{X},\mathbf{Y})],  \label{mc1}
\end{equation}%
for any $\mathbf{X},\ \mathbf{Y}\in \chi (M),$ and
\begin{equation}
\mathcal{D}(\sigma \beta )\doteq \mathcal{D}(\sigma )\beta +\sigma \mathcal{D%
}(\beta ),  \label{lc1}
\end{equation}%
for any $\sigma ,\beta \in Sec^{\infty }(\mathbf{E}).$

For local computations, we may define the corresponding coefficients of the
geometric d--objects and write
\begin{equation*}
\mathcal{D}\sigma _{\check{\beta}}\doteq {\mathbf{\Gamma }}_{\ \check{\beta}%
\mu }^{\check{\alpha}}\ \sigma _{\check{\alpha}}\otimes \delta u^{\mu }={%
\mathbf{\Gamma }}_{\ \check{\beta}i}^{\check{\alpha}}\ \sigma _{\check{\alpha%
}}\otimes dx^{i}+{\mathbf{\Gamma }}_{\ \check{\beta}a}^{\check{\alpha}}\
\sigma _{\check{\alpha}}\otimes \delta y^{a},
\end{equation*}%
where fiber ''inverse hat'' indices, in their turn, may split $\check{\alpha}%
\doteq ({\check{\imath}},{\check{a}})$ if any N--connection structure is
defined on $T\mathbf{E}.$ For some constructions of particular interest, we
can take $\mathbf{E}=T^{\ast }\mathbf{V},=T^{\ast }V_{(g)}$ and/or any
Clifford d--algebra $\mathbf{E}=\mathbb{C}l(\mathbf{V}),\mathbb{C}%
l(V_{(g)}),...$ with a corresponding treating of ''acute'' indices to of
d--tensor and/or d--spinor type as well when the d--operator $\mathcal{D}$
transforms into respective d--connection $\mathbf{D}$ and spin
d--connections $\widehat{\nabla }^{\mathbf{S}}$\ (\ref{csdc}), $\widehat{%
\nabla }^{(g)}$\ ... . All such, adapted to the N--connections, computations
are similar for both N--anholonomic (co) vector and spinor bundles.

The respective actions of the Clifford d--algebra and the
Clifford--Lagran\-ge algebra  can be transformed into maps
$Sec^{\infty }(\mathbf{S})\otimes Sec(\mathbb{C}l(\mathbf{V}))$ and \newline $%
Sec^{\infty }(S_{(g)})\otimes Sec(\mathbb{C}l(V_{(g)}))$ to $Sec^{\infty }(%
\mathbf{S})$ and, respectively, $Sec^{\infty }(S_{(g)})$ by considering maps
of type (\ref{gamfibb}) and (\ref{gamfibd})
\begin{equation*}
\widehat{\mathbf{c}}(\breve{\psi}\otimes \mathbf{a})\doteq \mathbf{c}(%
\mathbf{a})\breve{\psi}\mbox{\ and\ }\widehat{c}({\psi }\otimes {a})\doteq {c%
}({a}){\psi }.
\end{equation*}

\begin{definition}
\label{dddo} The Dirac d--operator (or Dirac N--anholonomic operator) on a
spin N--anholonomic manifold $(\mathbf{V},\mathbf{S},J)$ (or on a spin
manifold\newline
$(M_{(g)},S_{(g)},J))$ is defined
\begin{eqnarray}
\D &\doteq &-i\ (\widehat{\mathbf{c}}\circ \nabla ^{\mathbf{S}})  \label{ddo}
\\
&=&\left( \ ^{-}\D=-i\ (\ ^{-}\widehat{{c}}\circ \ ^{-}\nabla ^{\mathbf{S}%
}),\ ^{\star }\D=-i\ (\ ^{\star }\widehat{{c}}\circ \ ^{\star }\nabla ^{%
\mathbf{S}})\right)  \notag \\
(\ _{(g)}\D &\doteq &-i\ (\widehat{c}\circ \nabla ^{(g)})\ )  \label{dlo} \\
&=&\left( \ _{(g)}{^{-}\D=-i(\ ^{-}\widehat{c}\circ \ ^{-}\nabla ^{(g)}}),\
_{(g)}{^{\star }\D}=-i(\ ^{\star }\widehat{c}\circ \ ^{\star }\nabla
^{(g)})\right) \ ).  \notag
\end{eqnarray}%
Such N--adapted Dirac d--operators are called canonical and denoted $%
\widehat{\D}=(\ ^{-}\widehat{\D},\ ^{\star }\widehat{\D}\ )$\ ( $_{(g)}%
\widehat{\D}=(\ _{(g)}{^{-}\widehat{\D}},\ _{(g)}{^{\star }\widehat{\D}}\ )$%
\ ) if they are defined for the canonical d--connection and
respective spin d--connection (\ref{csdc}).
\end{definition}

Now we can formulate the

\begin{theorem}
\label{mr2} Let $(\mathbf{V},\mathbf{S},J)$ (\ $(M_{(g)},S_{(g)},J)$ be a
spin N--anholonomic manifold. There is the canonical Dirac d--operator
(Dirac N--anholonomic operator) defined by the almost Hermitian spin
d--operator
\begin{equation*}
\widehat{\nabla }^{\mathbf{S}}:\ Sec^{\infty }(\mathbf{S})\rightarrow
Sec^{\infty }(\mathbf{S})\otimes \Omega ^{1}(\mathbf{V})
\end{equation*}%
(N--anholonomic spin operator
\begin{equation*}
\widehat{\nabla }^{(g)}:\ Sec^{\infty }({S_{(g)}})\rightarrow Sec^{\infty
}(S_{(g)})\otimes \Omega ^{1}(M_{(g)})\ )
\end{equation*}%
commuting with $J$  and satisfying the conditions
\begin{equation}
(\widehat{\nabla }^{\mathbf{S}}\breve{\psi}\ |\ \breve{\phi})\ +(\breve{\psi}%
\ |\ \widehat{\nabla }^{\mathbf{S}}\breve{\phi})\ =\delta (\breve{\psi}\ |\
\breve{\phi})\   \label{scmdcond}
\end{equation}%
and
\begin{equation*}
\widehat{\nabla }^{\mathbf{S}}(\mathbf{c}(\mathbf{a})\breve{\psi})\ =\mathbf{%
c}(\widehat{\mathbf{D}}\mathbf{a})\breve{\psi}+\mathbf{c}(\mathbf{a})%
\widehat{\nabla }^{\mathbf{S}}\breve{\psi}
\end{equation*}%
for $\mathbf{a}\in \mathbb{C}l(\mathbf{V})$ and $\breve{\psi}\in Sec^{\infty
}(\mathbf{S})$
\begin{equation}
(\ (\widehat{\nabla }^{(g)}\breve{\psi}\ |\ \breve{\phi})\ +(\breve{\psi}\
|\ \widehat{\nabla }^{(g)}\breve{\phi})\ =\delta (\breve{\psi}\ |\ \breve{%
\phi})\   \label{scmlcond}
\end{equation}%
and
\begin{equation*}
\widehat{\nabla }^{(g)}(\mathbf{c}(\mathbf{a})\breve{\psi})\ =\mathbf{c}(%
\widehat{\mathbf{D}}\mathbf{a})\breve{\psi}+\mathbf{c}(\mathbf{a})\widehat{%
\nabla }^{(g)}\breve{\psi}
\end{equation*}%
for $\mathbf{a}\in \mathbb{C}l(M_{(g)})$ and $\breve{\psi}\in Sec^{\infty
}(S_{(g)}$ )\ determined by the metricity (\ref{mc1}) and Leibnitz (\ref{lc1}%
) conditions.
\end{theorem}

\begin{proof}
We sketch the main idea of such a proof being similar to that given in Ref. %
\cite{bondia}, Theorem 9.8, for the Levi--Civita connection, see also Ref. %
\cite{schroeder}. In our case, we have to extend the constructions for
d--metrics and canonical d--connections by applying N--elongated operators
for differentials and partial derivatives and distinguishing the formulas
into h-- and v--irreducible components. $\Box $
\end{proof}

The canonical Dirac d--operator has very similar properties for spin
N--anholonomic manifolds and spin Lagrange, or Finsler spaces \cite%
{vjmp,vhsp,vstav,vncl}.

\subsection{Field equations on N--anholonomic manifolds}

The general idea is to formulate such equations with respect to a
nonholonomic frame on (pesudo)\ Riemann--Cartan space. Then the
constructions are N--adapted by considering N--elongated frames. \ For
Lie/Clifford N--algebroid structures, we have to anchor the formulas.

\subsubsection{ Scalar field on N--anholonomic manifolds}

Let $\varphi \left( u\right) =(\varphi _{1}\left( u\right) ,\varphi
_{2}\left( u\right) \dot{,}...,\varphi _{k}\left( u\right) )$ be a complex
k-component scalar field of mass $\mu $ on a N--anholonomic manifold $%
\mathbf{V.}$ The d-covariant generalization of the conformally invariant (in
the massless case) scalar field equation \cite{penr1,penr2} can be defined
by using the d'Alambert operator $\square =\mathbf{D}^{\alpha }\mathbf{D}%
_{\alpha }$, where $\mathbf{D}_{\alpha }$ is a metric compatible
d--connection,
\begin{equation}
(\square +\frac{n+m-2}{4(n+m-1)}\overleftarrow{R}+\mu ^{2})\varphi \left(
u\right) =0.  \label{2.100}
\end{equation}%
We have to elongate the covariant d--operator, $\breve{D}_{\alpha }=\mathbf{D%
}_{\alpha }+ieA_{\alpha },$ and take into account the d-vector current
\begin{equation*}
J_{\alpha }^{(0)}\left( u\right) =i(\left( \overline{\varphi }\left(
u\right) \mathbf{D}_{\alpha }\varphi \left( u\right) -\mathbf{D}_{\alpha }%
\overline{\varphi }\left( u\right) )\varphi \left( u\right) \right)
\end{equation*}%
if there are considered interactions with the electromagnetic field (
d--vector potential $A_{\alpha }$), where $e$ is the electromagnetic
constant, and a charged scalar field $\varphi .$ The equations (\ref{2.100})
are just the Euler equations for the Lagrangian
\begin{equation}
\mathcal{L}^{(0)}\left( u\right) =\sqrt{|g|}\left[ \mathbf{g}^{\alpha \beta
}e_{\alpha }\overline{\varphi }\left( u\right) e_{\beta }\varphi \left(
u\right) -\left( \mu ^{2}+\frac{n+m-2}{4(n+m-1)}\right) \overline{\varphi }%
\left( u\right) \varphi \left( u\right) \right] ,  \label{2.101}
\end{equation}%
where $|g|=\det |\mathbf{g}_{\alpha \beta }|$ and $\mathbf{e}_{\alpha }$ is
defined by (\ref{dder}), and must be anchored for Lie algebroid structures.

The N--adapted variations of the action with Lagrangian (\ref{2.101}) on
variables $\varphi \left( u\right) $ and $\overline{\varphi }\left( u\right)
$ lead to the energy--momentum d--tensor,
\begin{equation}
E_{\alpha \beta }^{(0,c)}\left( u\right) =\mathbf{e}_{\alpha }\overline{%
\varphi }\left( u\right) \mathbf{e}_{\beta }\varphi \left( u\right) +\mathbf{%
e}_{\beta }\overline{\varphi }\left( u\right) \mathbf{e}_{\alpha }\varphi
\left( u\right) -\frac{1}{\sqrt{|g|}}\mathbf{g}_{\alpha \beta }\mathcal{L}%
^{(0)}\left( u\right) ,  \label{2.102}
\end{equation}%
and a similar variation on the components of a d--metric (\ref{metr}) leads
to a symmetric energy-momentum d-tensor,
\begin{equation}
E_{\alpha \beta }^{(0)}\left( u\right) =E_{(\alpha \beta )}^{(0,c)}\left(
u\right) -\frac{n+m-2}{2(n+m-1)}\left[ \mathbf{R}_{(\alpha \beta )}+\mathbf{D%
}_{(\alpha }\mathbf{D}_{\beta )}-\mathbf{g}_{\alpha \beta }\square \right]
\overline{\varphi }\left( u\right) \varphi \left( u\right) .  \label{2.103}
\end{equation}%
We also conclude that the N-connection results in a nonequivalence of
ener\-gy--momentum d-tensors (\ref{2.102}) and (\ref{2.103}), nonsymmetry of
the Ricci tensor, non--vanishing of the d-covariant derivation of the
Einstein d-tensor,\newline
$\mathbf{D}_{\alpha }\overleftarrow{\mathbf{G}}^{\alpha \beta }\neq 0$ and,
in consequence, a corresponding modification of conservation laws on
N--anholonomic manifolds.

\subsubsection{Proca equations}

Let consider a d-vector field $\varphi _{\alpha }\left( u\right) $ with mass
$\mu ^{2}$ (Proca field) interacting with exterior gravitational field. From
the Lagrangian
\begin{equation}
\mathcal{L}^{(1)}\left( u\right) =\sqrt{\left| \mathbf{g}\right| }\left[ -%
\frac{1}{2}{\overline{f}}_{\alpha \beta }\left( u\right) f^{\alpha \beta
}\left( u\right) +\mu ^{2}{\overline{\varphi }}_{\alpha }\left( u\right)
\varphi ^{\alpha }\left( u\right) \right] ,  \label{2.104}
\end{equation}%
where
\begin{equation*}
f_{\alpha \beta }=\mathbf{D}_{\alpha }\varphi _{\beta }-\mathbf{D}_{\beta
}\varphi _{\alpha },
\end{equation*}
one follows the Proca equations on N--anholonomic manifolds
\begin{equation}
\mathbf{D}_{\alpha }f^{\alpha \beta }\left( u\right) +\mu ^{2}\varphi
^{\beta }\left( u\right) =0.  \label{2.105}
\end{equation}%
Equations (\ref{2.105}) transform into a first type constraints for $\beta
=0.$ Acting with $\mathbf{D}_{\alpha }$ on (\ref{2.105}), for $\mu \neq 0$
we obtain second type constraints
\begin{equation}
\mathbf{D}_{\alpha }\varphi ^{\alpha }\left( u\right) =0.  \label{2.106}
\end{equation}

Putting (\ref{2.106}) into (\ref{2.105}) we obtain second order field
equations with respect to $\varphi _{\alpha }$ :
\begin{equation}
\square \varphi _{\alpha }\left( u\right) +\mathbf{R}_{\alpha \beta }\varphi
^{\beta }\left( u\right) +\mu ^{2}\varphi _{\alpha }\left( u\right) =0.
\label{2.107}
\end{equation}%
Anchoring of derivatives has to be considered for the operators $\mathbf{D}%
_{\alpha }$ and (as a consequence) for $\square $ and $\mathbf{R}_{\alpha
\beta }.$ The energy-momentum d-tensor and d-vector current following from
the (\ref{2.107}) can be written
\begin{equation*}
E_{\alpha \beta }^{(1)}\left( u\right) =-\mathbf{g}^{\varepsilon \tau
}\left( {\overline{f}}_{\beta \tau }f_{\alpha \varepsilon }+{\overline{f}}%
_{\alpha \varepsilon }f_{\beta \tau }\right) +\mu ^{2}\left( {\overline{%
\varphi }}_{\alpha }\varphi _{\beta }+{\overline{\varphi }}_{\beta }\varphi
_{\alpha }\right) -\frac{\mathbf{g}_{\alpha \beta }}{\sqrt{\left| \mathbf{g}%
\right| }}\mathcal{L}^{(1)}\left( u\right) .
\end{equation*}%
and
\begin{equation*}
J_{\alpha }^{\left( 1\right) }\left( u\right) =i\left( {\overline{f}}%
_{\alpha \beta }\left( u\right) \varphi ^{\beta }\left( u\right) -{\overline{%
\varphi }}^{\beta }\left( u\right) f_{\alpha \beta }\left( u\right) \right) .
\end{equation*}

For $\mu =0$ the d-tensor $f_{\alpha \beta }$ and the Lagrangian (\ref{2.104}%
) are invariant with respect to gauge transforms of type
\begin{equation*}
\varphi _{\alpha }\left( u\right) \rightarrow \varphi _{\alpha }\left(
u\right) +\delta _{\alpha }\Lambda \left( u\right) ,
\end{equation*}%
where $\Lambda \left( u\right) $ is a d-differentiable scalar function, and
we obtain a variant of Maxwell theory on N--anholonomic manifolds.

\subsubsection{Gravitons N--anholonomic backgrounds}

Let un consider a massless d--tensor field $\mathbf{q}_{\alpha \beta }\left(
u\right) $ as a small perturbation of the d--metric $\mathbf{g}_{\alpha
\beta }\left( u\right) .$ Considering, for simplicity, a torsionless
background we have the Fierz--Pauli equations
\begin{equation}
\square \mathbf{q}_{\alpha \beta }\left( u\right) +2\mathbf{R}_{\tau \alpha
\beta \nu }\left( u\right) ~\mathbf{q}^{\tau \nu }\left( u\right) =0
\label{2.108}
\end{equation}%
and d-gauge conditions
\begin{equation}
D_{\alpha }\mathbf{q}_{\beta }^{\alpha }\left( u\right) =0,\quad \mathbf{q}%
\left( u\right) \equiv \mathbf{q}_{\beta }^{\alpha }(u)=0,  \label{2.109}
\end{equation}%
where $\mathbf{R}_{\tau \alpha \beta \nu }\left( u\right) $ is curvature
d-tensor (these formulae can be obtained by using a perturbation formalism
with respect to $\mathbf{q}_{\alpha \beta }\left( u\right) ;$ in our case we
must take into account the distinguishing of geometrical objects.

We note that we can rewrite d-tensor formulas (\ref{2.100})--(\ref{2.109})
into similar d--spinor ones by considering spinor variables.

\subsubsection{N--anholonomic Dirac equation}

Let denote the Dirac d--spinor field by $\psi \left( u\right) =\left( \psi ^{%
\acute{\alpha}}\left( u\right) \right) $ and consider as the generalized
Lorentz transforms the group of automorphysm of the metric $g_{\widehat{%
\alpha }\widehat{\beta }}$ (for a N--adapted frame decomposition of
d-metric). The d--covariant derivation of field $\psi \left(
u\right) $ is written as
\begin{equation}
\overrightarrow{\bigtriangledown }_{\alpha }\psi =\left[ \mathbf{e}_{\alpha
}+\frac{1}{4}C_{\widehat{\alpha }\widehat{\beta }\widehat{\gamma }}\left(
u\right) ~l_{\alpha }^{\widehat{\alpha }}\left( u\right) \gamma ^{\widehat{%
\beta }}\gamma ^{\widehat{\gamma }}\right] \psi ,  \label{2.110}
\end{equation}%
where coefficients $C_{\widehat{\alpha }\widehat{\beta }\widehat{\gamma }%
}=\left( \mathbf{D}_{\gamma }l_{\widehat{\alpha }}^{\alpha }\right) l_{%
\widehat{\beta }\alpha }l_{\widehat{\gamma }}^{\gamma }$ generalize for
N--anholonomic spaces the corresponding Ricci coefficients on Riemannian
spaces. Using $\gamma $-objects $\gamma ^{\alpha }\left( u\right) $ (see (%
\ref{2.79})), we define the Dirac equations on
N--anholonomic manifolds:
\begin{equation}
(i\gamma ^{\alpha }\left( u\right) \overrightarrow{\bigtriangledown }%
_{\alpha }-\mu )\psi =0,  \label{2.111}
\end{equation}%
which are the Euler equations for the Lagrangian
\begin{eqnarray}
\mathcal{L}^{(1/2)}\left( u\right) &=&\sqrt{\left| g\right| }\{[\psi
^{+}\left( u\right) \gamma ^{\alpha }\left( u\right) \overrightarrow{%
\bigtriangledown }_{\alpha }\psi \left( u\right)  \label{2.112} \\
&&-(\overrightarrow{\bigtriangledown }_{\alpha }\psi ^{+}\left( u\right)
)\gamma ^{\alpha }\left( u\right) \psi \left( u\right) ]-\mu \psi ^{+}\left(
u\right) \psi \left( u\right) \},  \notag
\end{eqnarray}%
where $\psi ^{+}\left( u\right) $ is the complex conjugation and
transposition of the column $\psi \left( u\right) .$ We have to consider
anchoring of the operator $\overrightarrow{\bigtriangledown }_{\alpha }$ on
the N--anho\-lo\-no\-mic manifolds.

From (\ref{2.112}), we obtain the d-metric energy-momentum d-tensor
\begin{eqnarray*}
E_{\alpha \beta }^{(1/2)}\left( u\right) &=&\frac{i}{4}[\psi ^{+}\left(
u\right) \gamma _{\alpha }\left( u\right) \overrightarrow{\bigtriangledown }%
_{\beta }\psi \left( u\right) +\psi ^{+}\left( u\right) \gamma _{\beta
}\left( u\right) \overrightarrow{\bigtriangledown }_{\alpha }\psi \left(
u\right) \\
&&-(\overrightarrow{\bigtriangledown }_{\alpha }\psi ^{+}\left( u\right)
)\gamma _{\beta }\left( u\right) \psi \left( u\right) -(\overrightarrow{%
\bigtriangledown }_{\beta }\psi ^{+}\left( u\right) )\gamma _{\alpha }\left(
u\right) \psi \left( u\right) ]
\end{eqnarray*}%
and the d-vector source
\begin{equation*}
J_{\alpha }^{(1/2)}\left( u\right) =\psi ^{+}\left( u\right) \gamma _{\alpha
}\left( u\right) \psi \left( u\right) .
\end{equation*}%
We emphasize that interactions with exterior gauge fields can be introduced
by changing the locally anisotropic partial derivation from (\ref{2.110}) in
this manner:
\begin{equation}
e_{\alpha }\rightarrow e_{\alpha }+ie^{\star }B_{\alpha },  \label{2.113}
\end{equation}%
where $e^{\star }$ and $B_{\alpha }$ are respectively the constant and the
d-vector potential of gauge fields.

\subsubsection{Yang-Mills equations in d--spi\-nor form}

We consider a vector bundle $\mathcal{B}_{E},~\pi _{B}:\mathcal{B\rightarrow
}\mathbf{V}$ on $\mathbf{V.}$ Additionally to the d--tensor and d--spinor
indices, we use capital Greek letters, $\Phi ,\Upsilon ,\Xi ,\Psi ,...$ for
fibre (of this bundle) indices (see details in \cite{penr1,penr2}). Let $%
\underline{\bigtriangledown }_{\alpha }$ be, for simplicity, a torsionless,
linear connection in $\mathcal{B}_{E}$ satisfying conditions:
\begin{eqnarray*}
\underline{\bigtriangledown }_{\alpha } &:&{\ \Upsilon }^{\Theta
}\rightarrow {\ \Upsilon }_{\alpha }^{\Theta }\quad \left[ \mbox{or }{\ \Xi }%
^{\Theta }\rightarrow {\ \Xi }_{\alpha }^{\Theta }\right] , \\
\underline{\bigtriangledown }_{\alpha }\left( \lambda ^{\Theta }+\nu
^{\Theta }\right) &=&\underline{\bigtriangledown }_{\alpha }\lambda ^{\Theta
}+\underline{\bigtriangledown }_{\alpha }\nu ^{\Theta }, \\
\underline{\bigtriangledown }_{\alpha }~(f\lambda ^{\Theta }) &=&\lambda
^{\Theta }\underline{\bigtriangledown }_{\alpha }f+f\ \underline{%
\bigtriangledown }_{\alpha }\lambda ^{\Theta },\quad f\in {\ \Upsilon }%
^{\Theta }~[\mbox{or }{\ \Xi }^{\Theta }],
\end{eqnarray*}%
where by ${\ \Upsilon }^{\Theta }~\left( {\ \Xi }^{\Theta }\right) $ we
denote the module of sections of the real (complex) v-bundle $\mathcal{B}%
_{E} $ provided with the abstract index $\Theta .$ The curvature of
connection $\underline{\bigtriangledown }_{\alpha }$ is defined as
\begin{equation*}
K_{\alpha \beta \Omega }^{\qquad \Theta }\lambda ^{\Omega }=\left(
\underline{\bigtriangledown }_{\alpha }\underline{\bigtriangledown }_{\beta
}-\underline{\bigtriangledown }_{\beta }\underline{\bigtriangledown }%
_{\alpha }\right) \lambda ^{\Theta }.
\end{equation*}

For Yang-Mills fields, as a rule, one considers that $\mathcal{B}_{E}$ is
enabled with a unitary (complex) structure (complex conjugation changes
mutually the upper and lower Greek indices). It is useful to introduce
instead of $K_{\alpha \beta \Omega }^{\qquad \Theta }$ a Hermitian matrix $%
F_{\alpha \beta \Omega }^{\qquad \Theta }=i$ $K_{\alpha \beta \Omega
}^{\qquad \Theta }$ connected with components of the Yang-Mills d-vector
potential $B_{\alpha \Xi }^{\quad \Phi }$ according the formula:

\begin{equation}
\frac{1}{2}F_{\alpha \beta \Xi }^{\qquad \Phi }=\underline{\bigtriangledown }%
_{[\alpha }B_{\beta ]\Xi }^{\quad \Phi }-iB_{[\alpha |\Lambda |}^{\quad \Phi
}B_{\beta ]\Xi }^{\quad \Lambda },  \label{2.114}
\end{equation}%
where the spacetime indices commute with capital Greek indices. The gauge
transforms are written in the form:

\begin{eqnarray*}
B_{\alpha \Theta }^{\quad \Phi } &\mapsto &B_{\alpha \widehat{\Theta }%
}^{\quad \widehat{\Phi }}=B_{\alpha \Theta }^{\quad \Phi }~s_{\Phi }^{\quad
\widehat{\Phi }}~q_{\widehat{\Theta }}^{\quad \Theta }+is_{\Theta }^{\quad
\widehat{\Phi }}\underline{\bigtriangledown }_{\alpha }~q_{\widehat{\Theta }%
}^{\quad \Theta }, \\
F_{\alpha \beta \Xi }^{\qquad \Phi } &\mapsto &F_{\alpha \beta \widehat{\Xi }%
}^{\qquad \widehat{\Phi }}=F_{\alpha \beta \Xi }^{\qquad \Phi }s_{\Phi
}^{\quad \widehat{\Phi }}q_{\widehat{\Xi }}^{\quad \Xi },
\end{eqnarray*}%
where matrices $s_{\Phi }^{\quad \widehat{\Phi }}$ and $q_{\widehat{\Xi }%
}^{\quad \Xi }$ are mutually inverse (Hermitian conjugated in the unitary
case). The Yang-Mills d--equations are written
\begin{eqnarray}
\underline{\bigtriangledown }^{\alpha }F_{\alpha \beta \Theta }^{\qquad \Psi
} &=&J_{\beta \ \Theta }^{\qquad \Psi },  \label{2.115} \\
\underline{\bigtriangledown }_{[\alpha }F_{\beta \gamma ]\Theta }^{\qquad
\Xi } &=&0.  \label{2.116}
\end{eqnarray}%
We must introduce deformations of connection of type, $\bigtriangledown
_{\alpha }^{\star }~\longrightarrow \underline{\bigtriangledown }_{\alpha
}+P_{\alpha },$ (the deformation d-tensor $P_{\alpha }$ is induced by the
torsion in the vector bundle $\mathcal{B}_{E})$ into the definition of the
curvature of gauge fields (\ref{2.114}) and motion equations (\ref{2.115})
and (\ref{2.116}) if the interactions are considered for nontrivial torsions.

\section{Conclusions and Outlook}

In this work we formulated a spinor approach to the geometry of nonholonomic
spacetimes and classical field interactions with constraints  possessing Lie
algebroid symmetry. Such geometric constructions are performed for a special
case of nonholonomic distributions defining nonlinear connection
(N--connection) structures resulting in preferred classes of vielbein
(frame) systems of reference.  The main goals we have achieved are the
following:

\begin{enumerate}
\item We gave an intrinsic formulation of the geometry of Clifford
N--anholo\-no\-mic structures. In addition, we investigated the
N--anholo\-no\-mic spin structures (i. e. spinor nonholonomic spaces with
associated N--connection).

\item We defined and analyzed the main properties of the Dirac operator on
N--anholonomic manifolds. We showed how the formulas may be ''anchored'' in
order to be considered on spacetimes with Lie/ Clifford algebroid symmetries.

\item We formulated a geometric approach to field equations on
N--anholono\-mic manifolds. There were considered the examples of scalar,\
Proca, graviton, spinor and gauge filed interactions when the formulas have
a straightforward re--definition on Lie/Clifford N--algebroids (i. e.
spacetimes with algebroid symmetries and nonholonomic distributions).
\end{enumerate}

Among the subjects we will study in forthcoming papers, we note the points:

\begin{itemize}
\item To construct exact solutions of the gravitational field equations in
string gravity with nontrivial limits to general relativity, parametrized by
generic off--diagonal metrics and nonholonomic frames and possessing Lie
algebroid symmetries (the first examples of ''gravitational'' algebroids
were analyzed in Ref. \cite{valgexsol}).

\item Certain extension of the metrics to configurations defining solutions
of the Einstein--Dirac equations will be considered. We shall analyze the
symmetries of such spacetimes and possible physical applications in modern
gravity.

\item In explicit form, we shall construct nonholonomically deformed
metrics, with algebroid symmetries, describing locally anisotropic
cosmological models, black holes, anholonomic wormholes, solitons and
gravitational monopoles and instantons.

\item To make a detailed investigation of classical field theories and their
quantum deformations possessing nontrivial noncommutative symmetries and
possible Lie/ Clifford algebroid structure.
\end{itemize}

Finally, we note that the method of anholonomic frames with associated
N--connection structure elaborated in Finsler geometry and further, in our
works, applied to constructing exact solutions in gravity was applied in
this paper for a study of Dirac operators on nonholonomic manifolds
possessing Lie algebroid symmetry. The constructions can be extended for
spacetimes with uncompactified extra dimensions and such investigations are
regarded as interesting researches in modern physics and noncommutative
geometry.

\section*{Acknowledgments}

The author is grateful to the referee for hard work and constructive critics.

\end{document}